\documentclass{aa}  

\usepackage{graphicx}
\usepackage{txfonts}
\usepackage{amsmath,amssymb,dsfont}
\usepackage{xcolor}
\usepackage{hyperref}
\usepackage{comment}
\makeatletter
\renewcommand*\aa@pageof{, page \thepage{} of \pageref*{LastPage}}
\makeatother

\graphicspath{{./}{figures/}}
\newcommand{\bs}[1]{\boldsymbol{#1}}
\newcommand{\dust}{\ensuremath\mathrm{d}}
\newcommand{\gas}{\ensuremath\mathrm{g}}

\newcommand{\corr}[1]{#1}

\begin{document}
\title{$\mathcal{PT}$ and anti-$\mathcal{PT}$ symmetries for astrophysical waves}

   \subtitle{}

   \author{Armand Leclerc\inst{1}\fnmsep\thanks{armand.leclerc@ens-lyon.fr}
          \and
          Guillaume Laibe\inst{1}
          \and
          Nicolas Perez\inst{1}
          }

   \institute{Univ Lyon, Univ Lyon1, Ens de Lyon\\
            CNRS, Centre de Recherche Astrophysique de Lyon UMR5574\\
            F-69230, Saint-Genis,-Laval, France
            }

   \date{Received September 15, 1996; accepted March 16, 1997}

 
  \abstract
   {Discrete symmetries have found numerous applications in photonics and quantum mechanics, but remain little studied in fluid mechanics, particularly in astrophysics.}
   {We aim to show how $\mathcal{PT}$ and anti-$\mathcal{PT}$ symmetries determine the behaviour of linear perturbations in a wide class of astrophysical problems. They set the location of Exceptional Points in the parameter space and the associated transitions to instability, and are associated to the conservation of quadratic quantities that can be determined explicitly.}
   {We study several classical local problems: the gravitational instability of isothermal spheres and thin discs, the Schwarzschild instability, the Rayleigh-Bénard instability and acoustic waves in dust-gas mixtures. We calculate the locations and the order of the Exceptional Points with a method of resultant, as well as the conserved quantities in the different regions of the parameter space using Krein theory.}
   {All problems studied here exhibit discrete symmetries, even though Hermiticity is broken by different physical processes (self-gravity, buoyancy, diffusion, drag). This analysis provides genuine explanations for certain instabilities, and for the existence of regions in the parameter space where waves do not propagate. Those correspond to breaking of $\mathcal{PT}$ and anti-$\mathcal{PT}$ symmetries respectively. Not all instabilities are associated to symmetry breaking (e.g. the Rayleigh-Benard instability).}
   {}


\maketitle

\section{Introduction}

A large class of astrophysical systems, such as stars or discs, are often treated as fluids  (e.g. \citealt{pringle2007,armitage2010}). The properties and the stability of the corresponding linear modes are usually analysed with methods that, to our knowledge, do not fully exploit the symmetries of the perturbed system with respect to parity $\mathcal{P}$ and time $\mathcal{T}$.

Some of the most striking properties of $\mathcal{PT}$-symmetric systems emerged in the quantum physics community since the seminal paper by \citet{bender1998}, the first non-Hermitian quantum Hamiltonian situation with a real spectrum, which paved the way to open quantum systems. This framework has been extended to situations exhibiting a spontaneous $\mathcal{PT}$ symmetry breaking phase transition \citep{bender1999,bender2002}. Generalisations to quantum field theory have been developed in  \citet{bender2004} and considerations on the observables of such systems is found in \citet{mostafazadeh2004}. $\mathcal{PT}$ symmetry have been a particularly powerful tool in optics and photonics, where it has led to the development of novel designs with special properties. The analogy with quantum mechanics comes from the fact that beam dynamics is governed by a Schrödinger equation, where the optical index behaves as a complex potential, which can be designed to satisfy $\mathcal{PT}$ symmetry \citep{el-ganainy2007,makris2008}. \citet{klaiman2008} identified a spontaneous breaking of $\mathcal{PT}$ symmetry in a waveguide. \corr{\citet{musslimani2008} investigated the nonlinear propagation of optical solitons in $\mathcal{PT}$-symmetric refraction indexes.} \citet{guo2009,ruter2010} made the first observations of $\mathcal{PT}$ symmetry in optics with a complex index.  A decade of developments of $\mathcal{PT}$ symmetry in photonics lead to a multitude of applications (e.g. \citealt{feng2017,el-ganainy2018,ozdemir2019}). \citet{feng2014} use spontaneous breaking of $\mathcal{PT}$ symmetry to design a pure single-mode laser. \citet{peng2014} propose a design for a low-power optical diode, where the nonlinear regime of the $\mathcal{PT}$ symmetry induces non-reciprocal propagation. \citet{lin2011,regensburger2012,feng2013} use $\mathcal{PT}$ symmetry in periodic systems (through gratings, periodic crystals, or lattices) to design unidirectional invisibility. \corr{$\mathcal{PT}$-symmetric degeneracies are now used for their high resonant sensitivity in optical cavities \citep{hodaei2017}. Similar interest has been aroused by the use of anti-$\mathcal{PT}$ symmetry for novel optical designs (e.g. \citealt{ge2013,zhang2020synthetic,zhang2020breaking}).} The topological properties of $\mathcal{PT}$-symmetric systems have recently been investigated in the context of fluid dynamics \citep{fu2023}. $\mathcal{PT}$ and anti-$\mathcal{PT}$ symmetries have only recently been taken into account in fluid systems, to directly constrain fluid flows and instabilities by spontaneous symmetry breaking \citep{qin2019,fu2020d,david2022}. The $\mathcal{PT}$ symmetry framework has been demonstrated to be particularly adapted to tackle high-dimensional parameter space (e.g. the drift wave instabilities in tokamaks, \citealt{qin2021}). The richness underlying $\mathcal{PT}$-symmetric systems has thus been demonstrated unequivocally.

In this study, we aim to show that $\mathcal{PT}$ and anti-$\mathcal{PT}$ symmetries also control some of the properties of linear modes of astrophysical systems, such as instabilities or absence of propagation, stability exchange, conservation of quadratic quantities and this, for physical processes of different origins. We first define $\mathcal{PT}$ and anti-$\mathcal{PT}$ symmetries in Sect.~\ref{sec:def}. We then study how they condition the outcome of the linear analysis of four canonical astrophysical systems, namely the self-gravity instability in discs (Sect.~\ref{sec:toomre}, wave propagation in gas-dust mixtures (Sect.~\ref{sec:DW_pb}), and buoyancy instability in stellar (Sect.~\ref{sec:convection}) and planetary interiors (Sect.~\ref{sec:rayleighBenard}). We then discuss these analyses in a broader perspective for symmetric systems in Sect.~\ref{sec:discuss}.

\section{Definitions}
\label{sec:def}

\subsection{\texorpdfstring{$\mathcal{PT}$ symmetry}x}
\label{subsec:PT}
The evolution of a monochromatic linear perturbation to the steady state of a fluid can be written under the form of an eigenvalue problem ${H(a)X = \omega X}$, where the operator $H$ that describes the evolution of the perturbation depends continuously on some real parameter $a$. The complex eigenvalues are generically denoted by the frequency $\omega$, which can be a complex number \textit{a priori}. This problem is said to be $\mathcal{PT}$-symmetric if there exists a unitary operator $U$ such that 
\begin{equation}
    U^{}HU^{-1} = H^* \ ,
\label{eq:PT}
\end{equation} 
where $^*$ stands for complex conjugation. In this case, the eigenvalues go in pairs: if $\omega$ is an eigenvalue of the problem, so is $\omega^*$. Indeed, $HX = \omega X$ implies $H(U^{-1}X^*) = \omega^* (U^{-1}X^*)$. In $\mathcal{PT}$-symmetric systems, unstable modes are therefore accompanied by damped modes (imaginary parts of opposite sign). When $\omega \ne \omega^*$, $\mathcal{PT}$ symmetry is said to be {\it spontaneously broken} by $H$. By defining the symmetry operator $\mathcal{PT} \equiv U \Theta$ where $\Theta$ denotes complex conjugation, one has $[\mathcal{PT},H]=0$. The two operators commute, but do not share $X$ as an eigenvector for non-real eigenvalues.\\
\corr{\cite{mostafazadeh2002a,mostafazadeh2002b,zhang2020pt} showed that $\mathcal{PT}$ symmetry is equivalent to another symmetry, pseudo-Hermiticity, for finite-dimensional problems.} A problem described by the operator $H$ is pseudo-Hermitian if there exists an Hermitian operator $V$ such that
\begin{equation}
    VHV^{-1} = H^\dagger,
    \label{eq:psH}
\end{equation}
where ${}^\dagger$ denotes the Hermitian conjugate with respect to the appropriate Hermitian scalar product of the problem, which we will write $\cdot$ in the following such that $Y_1\cdot Y_2 \equiv Y_1^{*\top}Y_2$.\\
\corr{Let $X_1$ and $X_2$ be two eigenvectors of $H$ satisfying Eq.\eqref{eq:psH}. The eigenvalues $\omega_1$ and $\omega_2$ may take the same value: they degenerate. When the eigenvectors also become identical, the degeneracy is called an Exceptional Point (noted EP in the following). The linear stability of these systems is described by Krein theory \citep{krein1950,kirillov2013}: in such systems, the transition between stability and instability necessarily involves the exceptional degeneracy of stable eigenvalues. This phenomenon is known as a Krein collision. Before the collision, the system is stable, $\omega_1$ and $\omega_2$ are real, $\mathcal{PT}$ symmetry is said to be unbroken and the Krein quantities defined as $X_{1,2} \cdot V X_{1,2}$ are non-zero and of opposite signs. After the collision, the system is unstable, the complex frequencies $\omega_1 = \omega_2^*$ are complex conjugates of each other, and the two Krein quantities are necessarily zero.} The Krein quantity $Y\cdot V Y$
is a constant of motion of a general solution $Y(t,a)$ of the evolution equation $i\partial_t Y = H(a)Y$. Indeed,
\begin{eqnarray}
    i \partial_t (Y\cdot V Y) &=& Y\cdot (-H^\dagger V)Y + Y\cdot V HY\\
    &=& Y\cdot(VH-H^\dagger V)Y\\
    &=& 0.
\end{eqnarray}
Hence,
\begin{equation}
    Y\cdot V Y = \mathrm{cst}.
\end{equation}
Moreover, for a Fourier mode $Y = \mathrm{e}^ {-i\omega t}X$, this conserved quantity is exactly zero for unstable and damped modes, and non-zero for propagative modes. This is shown by expressing
\begin{align}
    &X\cdot V H X = X \cdot H^\dagger V X = \omega^* \;X \cdot V X\\
    &=\omega\; X\cdot V X.
\end{align}
According to the identity above, a situation where $\omega\neq \omega^*$ implies that $X\cdot V X = 0$.

\subsection{\texorpdfstring{Anti-$\mathcal{PT}$ symmetry}x}
\label{subsec:antiPT}
The eigenvalue problem $H \left(a \right)X = \omega X$ is said to be anti-$\mathcal{PT}$-symmetric, or alternatively $\mathcal{CP}$-symmetric, if there exists a unitary operator $\Tilde{U}$ such that 
\begin{equation}
    \Tilde{U}^{}H\Tilde{U}^{-1} = -H^*.
\label{eq:CP}
\end{equation}
In this case, the eigenvalues come in pairs: $\omega$ and $-\omega^*$. They have equal imaginary parts, and opposite real parts and describe counter-propagating waves. In particular, $H$ being anti-$\mathcal{PT}$-symmetric is equivalent for $iH$ to be $\mathcal{PT}$-symmetric. The same equivalence between $\mathcal{PT}$ symmetry and pseudo-Hermiticity then applies, and guarantees that an anti-$\mathcal{PT}$-symmetric operator $H$ is also pseudo-chiral: there exist an Hermitian operator $\Tilde{V}$ such that
\begin{equation}
    \Tilde{V}H\Tilde{V}^{-1} = -H^\dagger.
    \label{eq:psChi}
\end{equation}
The quantity $X\cdot \Tilde{V}X$ is not strictly a Krein quantity, and is not in general a constant of motion of a solution of $i\partial_t X = HX$. However, for a Fourier mode $Y = \mathrm{e}^ {-i\omega t}X$, it satisfies
\begin{equation}
    Y\cdot \Tilde{V} Y = \mathrm{e}^{2\mathrm{Im}(\omega)t} \; X\cdot \Tilde{V}X,
\end{equation}
and is constant to zero for modes with $\mathrm{Re}(\omega) \neq 0$, i.e for modes that spontaneously break the anti-$\mathcal{PT}$ symmetry. Indeed,
\begin{align}
    &X\cdot \Tilde{V} H X = X \cdot (-H^\dagger \Tilde{V}) X = -\omega^* \;X \cdot \Tilde{V} X\\
    &=\omega\; X\cdot \Tilde{V} X.
\end{align}
One concludes that whenever $\omega \neq -\omega^*$, $X\cdot \Tilde{V} X = 0$: the Krein quantity is zero at all times. An analogue of a Krein collision may occur. When the anti-$\mathcal{PT}$ symmetry is unbroken, their exist a pair of eigenvalues of $H$ denoted $\omega_1$ and $\omega_2$ which are purely imaginary. 

Consider the situation in which the parameter $a$ passes through an Exceptional Point, so that after the degeneracy we have $\omega_1=-\omega_2^*$ with non-zero real parts. Then, the anti-$\mathcal{PT}$ symmetry is spontaneously broken. This phenomenon corresponds to a Krein collision for the operator $iH$. It follows that the quantity $X\cdot \Tilde{V}X$ is non-zero and not conserved for non-propagative modes, and strictly zero for propagating modes for all times. \\
Systems that are both $\mathcal{PT}$ and anti-$\mathcal{PT}$ symmetric are called \textit{bi-symmetric}.

\corr{The eigenvalues of bi-symmetric systems take the form of $(\omega,-\omega)$ pairs of real or purely imaginary frequencies. When $\omega$ is real, the associated eigenvector spontaneously breaks the anti-$\mathcal{PT}$ symmetry. Conversely, when $\omega$ is imaginary, the associated eigenvector spontaneously breaks $\mathcal{PT}$ symmetry. Thus, one of those two symmetries is necessarily broken for any value of the parameters $a$, and either $X\cdot V X = 0$ (broken $\mathcal{PT}$) or $X\cdot \Tilde{V} X = 0$ (broken anti-$\mathcal{PT}$). In bi-symmetric systems, $X\cdot V X$ is always a constant of motion, while $X\cdot \Tilde{V} X$ is constant only when it is zero, in the spontaneously broken anti-$\mathcal{PT}$ symmetry phase.}

\section{Gravitational instabilities}
\label{sec:toomre}

\subsection{Jeans instability}
 The stability of self-gravitating objects plays a central role in the formation of structures throughout the Universe. The simplest problem to study would be the Jeans instability of a collapsing sphere \citep{jeans1902}. We consider local radial adiabatic perturbations of a 3D homogeneous self-gravitating, non-rotating, inviscid sphere of constant density $\rho_{0}$ and sound speed $c_{\rm s}$. Linear perturbations of mass and momentum conservation as well as Poisson equation form a 4$\times$4 system that can be written after Fourier transforms in time and space
\begin{equation} \label{eq:Jeans_eigenvalue}
    \omega \begin{pmatrix}
        \rho'/\rho_0 \\[5pt] \Vec{v}'/c_\mathrm{s}
    \end{pmatrix}
    = H_\mathrm{J}\begin{pmatrix}
        \rho'/\rho_0 \\[5pt] \Vec{v}'/c_\mathrm{s}
    \end{pmatrix}
    \equiv \begin{pmatrix}
        0 & c_\mathrm{s} \Vec{k}^\top \\[5pt]
        \left(1-\frac{4\pi\mathcal{G}\rho_0}{c_\mathrm{s}^2\vert k\vert^2}\right)c_\mathrm{s} \Vec{k} & 0 
    \end{pmatrix}
    \begin{pmatrix}
        \rho'/\rho_0 \\[5pt] \Vec{v}'/c_\mathrm{s}
    \end{pmatrix}.
\end{equation}
This problem is invariant under both parity and time reversal. It is therefore trivially $\mathcal{PT}$-symmetric, as can be proved by the statement $H_\mathrm{J} = H_\mathrm{J}^*$. As such, $H_\mathrm{J}$ is pseudo-Hermitian, with the symmetry operator $V = \mathrm{diag}\left(\left(1-\frac{4\pi\mathcal{G}\rho_0}{c_\mathrm{s}^2\vert k\vert^2}\right),1,1,1\right).$ The Krein quantity associated to the Krein collision of the Jeans instability is then $X\cdot V X = \left(1-\frac{4\pi\mathcal{G}\rho_0}{c_\mathrm{s}^2\vert k\vert^2}\right)\vert \rho^{\prime}\vert^2 + \vert \Vec{v}'\vert^2$. This quantity is non-zero for propagating waves, and zero for the unstable modes. Complementarily, $H_\mathrm{J}$ is anti-$\mathcal{PT}$ symmetric and pseudo-chiral. Indeed, it satisfies Eq.\ref{eq:psChi} for $\Tilde{V} = \mathrm{diag}(-\left(1-\frac{4\pi\mathcal{G}\rho_0}{c_\mathrm{s}^2\vert k\vert^2}\right),1,1,1)$. The quantity $X\cdot \Tilde{V} X = -\left(1-\frac{4\pi\mathcal{G}\rho_0}{c_\mathrm{s}^2\vert k\vert^2}\right)\vert \rho^{\prime}\vert^2 + \vert \Vec{v}'\vert^2$ is then zero for propagative modes and non-constant and non-zero for  unstable modes. The Krein collision occurs at an Exceptional Point for which $H$ cannot be diagonalised. This EP corresponds to marginal stability, which is reached at the Jeans wavenumber $\vert\Vec{k}\vert=k_\mathrm{J}\equiv \sqrt{4\pi\mathcal{G}\rho_0}/c_\mathrm{s}$. Longer wavelengths are unstable, and shorter wavelengths are propagating.

\subsection{Toomre instability}

 More interesting is the case of a self-gravitating disc \citep{goodman1988,goodman2003,bertin1999}. In this system, $\mathcal{P}$ and $\mathcal{T}$ are broken individually, but the overall system of linear perturbations is still $\mathcal{PT}$-symmetric. Stabilization of large and small scales by rotation and pressure can be sufficient to stabilize astrophysical discs against gravitational collapse \citep{toomre1964}.  We discuss here the simplest case of axisymmetric perturbations of short radial wavelengths evolving in a razor-thin Keplerian disc. The stability of linear perturbations is given by the celebrated Toomre criterion: $Q \equiv \frac{c_\mathrm{s}\kappa}{\pi\mathcal{G}\Sigma_0}>1$, where $\kappa$ denotes the epicyclic frequency of the disc and $\Sigma_0$ its surface density. The razor-thin disc is integrated vertically \citep{armitage2010}, to give the following set of radial linear perturbation

\begin{eqnarray}
    H_\mathrm{sgd}\begin{pmatrix}
        h'\\ v_r' \\v_\phi'
    \end{pmatrix}=\left(
\begin{array}{ccc}
 0 & c_\mathrm{s} k & 0 \\
 -\frac{k}{c_\mathrm{s}} \left(\frac{2}{Q|k| }-1\right) & 0 & 2 i \\
 0 & -\frac{i}{2} & 0 \\
\end{array}
\right)\begin{pmatrix}
        h'\\ v_r' \\v_\phi'
    \end{pmatrix}=\omega \begin{pmatrix}
        h'\\ v_r' \\v_\phi'
    \end{pmatrix}, \nonumber\\
\end{eqnarray}
where we used perturbations of the enthalpy $h'$ and the two horizontal velocities $v_r'$ and $v_{\phi}'$ as variables, and the orbital time $\Omega^{-1}$ and the pressure length $c_\mathrm{s} \Omega^{-1}$ as units of time and length.

\begin{figure}
    \centering
    \includegraphics[width=\columnwidth]{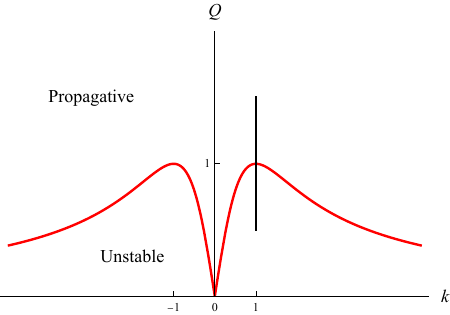}\\
    \includegraphics[width=\columnwidth]{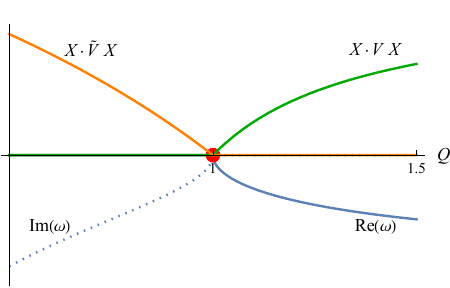}
    \caption{Top panel : curve of Exceptional Points corresponding to marginal stability. Maximum is reached for $Q = 1$ as expected. Bottom panel : Krein collision of the stability of the Toomre problem when varying the parameters along the black line of the top panel. Depending on the value of $Q$, exactly one of the two Krein quantities identified is zero, depending on which symmetry is broken by the perturbation. \corr{The curve in green is $X\cdot VX$, orange is $X\cdot\Tilde{V}X$, solid blue is $\mathrm{Re}(\omega)$ and dotted blue is $\mathrm{Im}(\omega)$, for a destabilizing mode.}}
    \label{fig:toomreCollision}
\end{figure}

With $U = \mathrm{diag}(1,1,-1)$, this problem satisfies Eq.~\eqref{eq:PT} and is therefore $\mathcal{PT}$-symmetric. Rotation breaks the reflection symmetry in the azimuthal direction and the time reversal symmetry, but not the combination of both. It is expected that the system is pseudo-Hermitian, which can be shown explicitly. With
\begin{equation}
    V = \left(
\begin{array}{ccc}
 \frac{1}{c_\mathrm{s}^2} & 0 & -\frac{4 i k}{c_\mathrm{s}Q | k| } \\
 0 & 1 & 0 \\
 \frac{4 i k}{c_\mathrm{s}Q | k| } & 0 & 4 \left(\frac{2 | k| }{Q}+1\right) \\
\end{array}
\right) ,
\label{eq:VToomre}
\end{equation}
the system satisfies Eq.~\eqref{eq:psH}. The Krein quantity associated with this $\mathcal{PT}$ symmetry provides the following energy partition, which applies to the unstable mode
\begin{equation}
\begin{split}
    X\cdot VX = 0 = \frac{\vert h'\vert^2}{c_\mathrm{s}^2} + \vert v_r'\vert^2 &+ 4\vert v_\phi'\vert^2\left(1+\frac{2\pi\mathcal{G}\Sigma_0}{\Omega^2}\vert k\vert\right) \\ 
    &- \frac{8\pi\mathcal{G}\Sigma_0}{c_\mathrm{s}^2\Omega}\text{sgn}(k)\mathrm{Im}\left(h v_{\phi }^*\right),
\end{split}
\end{equation}
where $k$ denotes the wavenumber in dimensional form. We illustrate this result by computing numerically the value of this quantity through a Krein collision, since $H_{\rm sgd}$ has degenerated eigenvalues when $Q = \frac{2\vert k\vert}{1+k^2}$. These are the positions of Exceptional Points for $H_\mathrm{sgd}$, which corresponds to positions of a Krein collision. The curve of Exceptional Points separates two regions in $(Q,k)$- space: one is the stable region, the other is the unstable region.
Figure~\ref{fig:toomreCollision} shows the EPs curve and the Krein collision that occurs when crossing this curve.

In addition to being $\mathcal{PT}$-symmetric, $H_{\rm sgd}$ is also anti-$\mathcal{PT}$-symmetric since
\begin{eqnarray}
    \Tilde{U} H_{\rm sgd} \Tilde{U}^{-1} = -H_{\rm sgd}^*,
\end{eqnarray}
with $\Tilde{U} = \mathrm{diag}(1,-1,-1)$. The anti-$\mathcal{PT}$ symmetry is complementary to the $\mathcal{PT}$ symmetry, and the combination of the two symmetries implies that the eigenvalues always come in pairs: they are always opposite $(\omega,-\omega)$, either real or purely imaginary. They are never general complex numbers, a generic property for bi-symmetric systems, since one of the two symmetries is necessarily spontaneously broken.\\
One has
\begin{eqnarray}
    \Tilde{V} H_\mathrm{sgd} \Tilde{V}^{-1} &=& -H_\mathrm{sgd}^\dagger,\\
    \Tilde{V} &\equiv& \left(
\begin{array}{ccc}
 \frac{1}{c^2} & \;\;0\;\; & -\frac{4 i k (Q | k| -1)}{c Q | k| } \\
 0 & \;\;1\;\; & 0 \\
 \frac{4 i k (Q | k| -1)}{c Q | k| } &  \;\;0\;\; & -\frac{8 k^2}{Q | k| }+8 k^2-4 \\
\end{array}
\right).
\label{eq:Toomre_pc}
\end{eqnarray}
Equation~\eqref{eq:Toomre_pc} explicitly shows that $H_\mathrm{sgd}$ is pseudo-chiral as expected, and Krein theory then applies. When $iH_\mathrm{sgd}$ has real eigenvalues, anti-$\mathcal{PT}$ symmetry is unbroken, and the Krein quantities $X \cdot \Tilde{V} X$ are non-zero. This holds when $\mathcal{PT}$-symmetry is spontaneously broken, i.e in the unstable region of parameters, where the first Krein quantity $X \cdot V X$ is zero. Conversely, in the stable range, anti-$\mathcal{PT}$ symmetry is broken and $\mathcal{PT}$ symmetry is unbroken. In this case, $X \cdot \Tilde{V} X$ is zero and $X \cdot V X$ is not equal to zero. The Krein quantity $X \cdot \Tilde{V} X$ associated to the anti-$\mathcal{PT}$-symmetry gives the following energy distribution for propagating waves in dimensional form
\begin{equation}
\begin{split}
    X\cdot \Tilde{V}X = &\frac{\vert h'\vert^2}{c_\mathrm{s}^2} + \vert v_r'\vert^2 + \vert v_\phi'\vert^2\left(-4+8\left(\frac{c_\mathrm{s}^2 k^2}{\Omega^2}-\frac{\pi \mathcal{G}\Sigma_0}{\Omega^2}\vert k\vert\right)\right) \\
    &- \frac{8\pi\mathcal{G}\Sigma_0}{c_\mathrm{s}^2\Omega}\text{sgn}(k)\left(\frac{c_\mathrm{s}^2}{\pi\mathcal{G}\Sigma_0}\vert k\vert -1\right)\mathrm{Im}\left(h v_{\phi }^*\right) .
\end{split}
\end{equation}

Hence, for each mode, either $X \cdot V X$ or $X \cdot \Tilde{V} X$ is necessarily zero: the former is zero for unstable modes and the latter is zero for propagating waves. Both are exactly zero at the degeneracy, i.e. for $\omega=0$, which is true for $Q = \frac{2\vert k\vert}{1+k^2}$. Figure~\ref{fig:toomreCollision} shows this exchange in non-zero quantities when crossing the Krein collision. Krein quantities simply provide energy partitions of the system in both regimes.

The matrices $V$ and $\tilde{V}$ identified above are particular cases of more general families of matrices
\begin{equation}
    V = \left(
\begin{array}{ccc}
\frac{2 c_{\rm s} \left( - 2k^{2} + \left(-1 + k^{2} \right) Q \left| k \right| \right)a + k\left( -2 + Q\left| k \right|\right)c }{4 c_{\rm s}^{2} k Q \left| k \right|} & b \frac{\left( kQ - 2 \mathrm{sgn{k}}\right)}{2 c_{\rm s} Q} & i a \\
 b \frac{\left( kQ - 2 \mathrm{sgn{k}}\right)}{2 c_{\rm s} Q} & \frac{1}{4} \left(2 c_{\rm s} k a + c \right) & i b \\
 -ia & -ib & c \\
\end{array}
\right),
\label{eq:VToomre_gene}
\end{equation}
with $a,b,c$ any real numbers such that $\det W \neq 0$. The matrix of Eq.~\eqref{eq:VToomre} is recovered with $a = -\frac{4k}{c_{\rm s}Q\left| k\right|}$, $b = 0$ and $c = \frac{8 \left| k\right|}{Q} + 2 $. Similarly, %
\begin{equation}
    \tilde{V} = \left(
\begin{array}{ccc}
\frac{2 c_{\rm s} \left( - 2k^{2} + \left(-1 + k^{2} \right) Q \left| k \right| \right)a + k\left( -2 + Q\left| k \right|\right)c }{4 c_{\rm s}^{2} k Q \left| k \right|} & b \frac{i\left( kQ - 2 \mathrm{sgn{k}}\right)}{2 c_{\rm s} Q} & i a \\
- b \frac{i\left( kQ - 2 \mathrm{sgn{k}}\right)}{2 c_{\rm s} Q} & - \frac{1}{4} \left(2 c_{\rm s} k a + c \right) & b \\
 -i a & b & c \\
\end{array}
\right)
\label{eq:tildeVToomre_gene}
\end{equation}
The matrix of Eq.~\eqref{eq:Toomre_pc} is recovered with $a = -\frac{4k(Q\vert k\vert-1)}{c_\mathrm{s}Qk}$, $b = 0$ and $c = -\frac{8}{Q}\vert k\vert+8k^2-4$. The same remark holds for the different problems adressed in this study. 

\section{Buoyancy instability in stars}
\label{sec:convection}
 Consider a non-rotating star in static equilibrium, balanced by thermal pressure and self-gravity. The steady state is spherically symmetric, the density is stratified and decreases towards the surface. This equilibrium can be unstable, since a perturbation can be amplified by buoyancy if the square of the buoyancy frequency $N^2$ is negative, as given by the Schwarzschild criterion \citep{schwarzschild1906}. 
 
\begin{figure}
    \centering
    \includegraphics[width=\columnwidth]{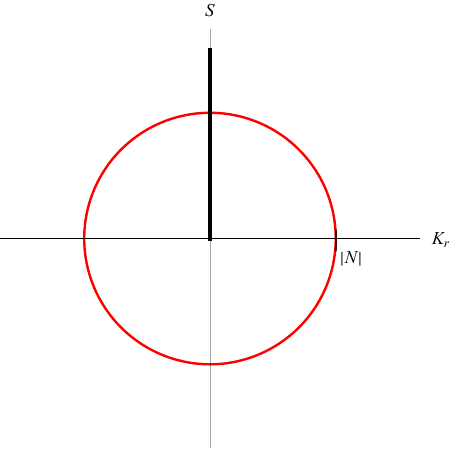}\\
    \includegraphics[width=\columnwidth]{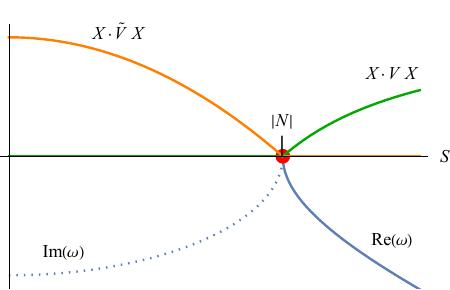}
    \caption{Same as Fig.~\ref{fig:toomreCollision} for the Schwarzschild instability. Non-hermiticity is broken in this case by buoyancy instead of self-gravity.}
    \label{fig:conv_collision}
\end{figure}

This problem has been revisited by \citet{leclerc2023} in inhomogeneous media, where the system proved to be pseudo-Hermitian and pseudo-chiral. Pseudo-Hermiticity was used to determine the Krein quantity. We will now complete the picture by determining the Krein quantity associated to pseudo-chirality, and show the Krein collision in this bi-symmetric problem.\\
Neglecting the self-gravity of the perturbations (Cowling's approximation \citealt{cowling1941}), we start from Eq.5 of \citet{leclerc2023}, which is the eigenvalue equation
\begin{eqnarray}
    \omega X &=& H_{\rm b} X,\\
    H_{\rm b} &=& \begin{pmatrix}
    0 & 0 & 0 & L_\ell \\
    0 & 0 & \vert N \vert & K_r - iS \\
    0 & - \vert N \vert & 0 & 0 \\
    L_\ell & K_r +iS & 0 & 0
\end{pmatrix}
\end{eqnarray}
where $\omega$ is the complex eigenfrequency of the perturbation, and $X \equiv  \begin{pmatrix} {v} & {w} & {\Theta} & {p} \end{pmatrix}^\top$ contains the perturbation's horizontal velocity, radial velocity, entropy and pressure after appropriate rescaling. $N^2 \equiv -g\frac{\mathrm{d}{\ln\rho_0}}{\mathrm{d}r} - \frac{g^2}{c_\mathrm{s}^2}$ is the square of the buoyancy frequency which is negative here, $S \equiv \frac{c_\mathrm{s}}{2g}\left( N^2 - \frac{g^2}{c_\mathrm{s}^2} \right) - \frac{1}{2}\frac{\mathrm{d}{c_\mathrm{s}}}{\mathrm{d}r} + \frac{c_\mathrm{s}}{r}$ is another characteristic frequency called the buoyant-acoustic frequency, which  quantifies the coupling between $g$-modes and $p$-modes in asteroseismic problems, $L_\ell = \frac{c_\mathrm{s}}{r}\sqrt{\ell(\ell+1)}$ is the Lamb frequency, $K_r$ represents the local radial wavenumber of the wave (see Appendix E of \cite{leclerc2023} for details).\\
This problem is both pseudo-Hermitian and pseudo-chiral, and is  associated with the two matrices
\begin{eqnarray}
    V &=& \mathrm{diag}(1,\;\;\;1,-1,\;\;\;1),\\
    \Tilde{V} &=& \mathrm{diag}(1,\;\;\;1,\;\;\;1,-1).  
\end{eqnarray}
Krein theory can therefore be applied to determine how discrete symmetries constrain the partition of energy. $X \cdot X$ is generally not a conserved quantity, since $H_{\rm b}$ is not Hermitian. It grows exponentially for unstable modes and is conserved only for stable modes. However, $X \cdot V X$ is a constant of motion for any solution of $\partial_t X = H_{\rm b}X$, as it has been shown in Sect.\ref{subsec:PT}. Moreover, this constant is exactly zero only for unstable modes. In contrast, $X \cdot \Tilde{V} X$ also grows exponentially for unstable modes, but is constantly equal to zero for stable modes. As such,
\begin{align}
    X &\cdot X &\hspace{-1.5cm}= \vert v\vert^2 + \vert w\vert^2 + \vert \Theta\vert^2 + \vert p\vert^2 &\propto \mathrm{e}^{2 \mathrm{Im}(\omega)t}, \\
    X &\cdot VX &\hspace{-1.5cm}=  \vert v\vert^2 + \vert w\vert^2 - \vert \Theta\vert^2 + \vert p\vert^2 &=\mathrm{cst}\; \delta_{\mathrm{Im}(\omega),0}, \\
    X &\cdot \Tilde{V} X &\hspace{-1.5cm}= \vert v\vert^2 + \vert w\vert^2 + \vert \Theta\vert^2 - \vert p\vert^2 &\propto \mathrm{e}^{2 \mathrm{Im}(\omega)t} \delta_{\mathrm{Re}(\omega),0},
\end{align}
where $\delta$ is the Kronecker delta. \\
For the radial modes $L_\ell=0$, the modes degenerate over the ring parameterised by $K_r^2+S^2 = -N^2 > 0$. These parameter values are Exceptional Points, as $H_{\rm b}$ can no longer be diagonalized there. The ring separates two ranges of parameters, the stable and the unstable part. Figure~\ref{fig:conv_collision} shows the Krein quantities, over the ring of EPs in which the collision takes place. The Krein collision behaves as expected: inside the ring, the system is unstable, and $X\cdot VX$ is zero and $X\cdot \Tilde{V}X$ is non-zero. Outside the ring, the modes are stable and propagate as sound waves, and $X\cdot VX$ is non-zero and $X\cdot \Tilde{V}X$ is zero. 

This structure is equivalent to the bi-symmetric problem of self-gravitating discs discussed in Sect.\ref{sec:toomre}, and shown on Fig.\ref{fig:toomreCollision}, even if the physical situation is very different. The approach followed in this problem is the reverse of that followed in section ~\ref{sec:toomre} for the Toomre problem: it is easier to first establish that the system is pseudo-Hermitian and pseudo-chiral, since the matrices $V$ and $\Tilde{V}$ have $\pm 1$ on the diagonal.

\section{Rayleigh-Bénard instability}
\label{sec:rayleighBenard}
\begin{figure}
    \centering
    \includegraphics[width=\columnwidth]{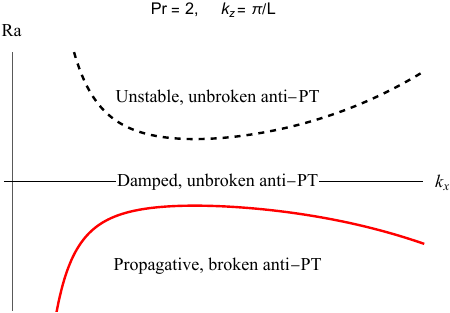}
    \caption{Different regions of the Rayleigh-Bénard problem in \corr{the parameter} space. In the upper part $\mathrm{Ra}>0$, anti-$\mathcal{PT}$ symmetry is protected, frequencies are purely imaginary, the system exhibits an instability (black dashed) and no EPs. In the lower part $\mathrm{Ra}<0$, anti-$\mathcal{PT}$ is no longer protected by anti-Hermiticity, and the system exhibits a curve of EP2s (red), delimiting a region where perturbations propagate. It is a region where anti-$\mathcal{PT}$ symmetry is spontaneously broken.}
    \label{fig:rayleigh-benard}
\end{figure}

Not all instabilities correspond to spontaneous symmetry breakings. The Rayleigh-Bénard instability \citep{benard1900,rayleigh1916} is such a case. Interiors of rocky planets are often treated as incompressible with both significant viscous and thermal diffusion (e.g.  \citealt{berge1984,bodenschatz2000,brandenburg2021}). This results in a regime of buoyancy instability that is controlled by diffusion effects and differs fundamentally from that discussed in Sect.~\ref{sec:convection} for adiabatic perturbations in compressible stars or planetary atmospheres. In the Rayleigh-Bénard regime, the instability criterion is given by the value of the Rayleigh number $\mathrm{Ra} = g\alpha\beta L^4 /\nu\kappa$, which needs to be greater than some critical value of order $10^{2} - 10^{3}$ which depends on the boundary conditions. This famous problem admits analytical solutions for an homogeneous background configuration with rigid-lids isothermal boundary conditions, which provides an illustrative example in this study for discussing the symmetries involving diffusive effects.\\
The basic set of equations consists of the incompressibility condition, a Navier-Stokes equation, and thermal heat diffusion for a 2D $(x,z)$ fluid (see Appendix \ref{app:rayleigh-benard}). The $x$- direction is invariant by translation and infinite, the $z$- direction is of length $L$, along which a constant temperature gradient $\partial_z T_0 = -\beta$ is directed. With dimensionless variables, the problem for Fourier modes $\exp{(-i\omega t + ik_z z + ik_x x)}$ can be converted into the symmetrical form
\begin{align}
    &\omega \begin{pmatrix}
        \Tilde{v_z}\\[0.2cm] \Tilde{T}
    \end{pmatrix} = H_{\rm RB} \begin{pmatrix}
        \Tilde{v_z}\\[0.2cm] \Tilde{T}
    \end{pmatrix},\\
    &H_{\rm RB} = -i\begin{pmatrix}
        k^2 & -\frac{k_x}{k}\left(\frac{\mathrm{Ra}}{\mathrm{Pr}}\right)^{1/2}\\
        -\frac{k_x}{k}\left(\frac{\mathrm{Ra}}{\mathrm{Pr}}\right)^{1/2} & k^2/\mathrm{Pr}
    \end{pmatrix},
    \label{eq:rayleigh-benard-operator}
\end{align}
where $\mathrm{Pr}=\nu/\kappa$ is the Prandtl number and $k = \sqrt{k_x^2+k_z^2}$.\\
This problem is anti-$\mathcal{PT}$-symmetric and pseudo-chiral, since one reads that
\begin{equation}
    H_{\rm RB} = -H_{\rm RB}^* = -H_{\rm RB}^\dagger,
\end{equation}
for $\mathrm{Ra} \geq 0$. In this case, the anti-$\mathcal{PT}$ symmetry is protected and cannot be spontaneously broken since $H_{\rm RB}$ is anti-Hermitian (i.e. $iH_{\rm RB}$ is Hermitian). Indeed, the spectral theorem guarantees that the eigenvalues are purely imaginary and the eigenvectors are orthogonal to each other.\\
The location of EPs is given by
\begin{equation}
    \mathrm{Ra} = \left(1-\frac{(1+\mathrm{Pr})^2}{\mathrm{4 Pr}}\right)\frac{(k_x^2 + k_z^2)^3}{k_x^2},
\end{equation}
which requires $\mathrm{Ra}$ to be negative. This condition can only be fulfilled for an inverted temperature gradient, which corresponds to a stably stratified liquid. In this case, the anti-$\mathcal{PT}$ symmetry is no longer protected and can be broken spontaneously, which would lead to propagating, damped internal waves.\\
\corr{Anti-Hermiticity ensures that the eigenvalues $\omega = i\eta$ are purely imaginary for all $k_x,k_z,\mathrm{Ra},\mathrm{Pr}$. The transition to instability is given by the change of sign $\eta < 0$ to $\eta >0$. This necessarily occurs for $\omega = 0$, providing the marginal stability criterion for the Rayleigh-Bénard instability.} This condition is \corr{satisfied when}
\begin{equation}
    \mathrm{Ra} = \frac{(k_x^2 + k_z^2)^3}{k_x^2},
\end{equation}
whose minimum for each $k_x$,$k_z$ compatible with the boundary conditions gives the critical value of the Rayleigh number (e.g. $\mathrm{Ra_{c}} = \frac{27}{4}\pi^4$ for stress-free, impenetrable and isothermal boundaries). Figure~\ref{fig:rayleigh-benard} shows the regions of this problem.

\section{Waves in dusty mixtures}
\label{sec:DW_pb}

Mixtures of pressureless dust and inviscid non-magnetised gas are generically used to model basic hydrodynamical properties of the dusty interstellar medium \corr{(e.g. \citealt{saffman1962,baines1965,ahuja1973,gumerov1988}}). Exchange of momentum between gas and dust is modeled as a drag force proportional to the difference in their velocities. The \textsc{dustywave} problem consist of studying the propagation of a 1D acoustic wave in an homogeneous dusty medium. The question is whether small disturbances propagate or are simply attenuated, since both regimes are possible  (e.g.  \citealt{laibe2011,david2021}). The mixture is initially homogeneous and at rest. Looking for small perturbative solutions as Fourier modes $\exp{(i(\omega t - k x))}$, one obtains the eigenvalue equation
\begin{align} 
    &H_{\rm dw} X = \omega X,\\
    &H_{\rm dw} \equiv \begin{pmatrix}
    0 & \;\;\; 0 &\;\;\;\; k & 0 \\
    0 &\;\;\; 0 &\;\;\;\; 0 & k \\
    k &\;\;\; 0 &\;\;\;\; i\epsilon  &  \;-i(1-\epsilon) \\
    0 &\;\;\; 0 &\; -i\epsilon  & \;\;\;i(1-\epsilon)
    \end{pmatrix},
    \label{eq:eigen_dusty}
\end{align}
after appropriate rescaling and choice of dimensionless parameters given in App.~\ref{app:dustyEq}.
The operator $H_{\rm dw}$ formally depends  on two parameters: the wavenumber $k$ and the dust density fraction of the mixture $\epsilon$. $H_{\rm dw}$ satisfies Eq.~\eqref{eq:CP} with $\Tilde{U} = \mathrm{diag}(1,1,-1,-1)$, meaning that $H_{\rm dw}$ is anti-$\mathcal{PT}$-symmetric. This property stems from the reflection symmetry $x\mapsto -x$ of the fluid in physical space, as it has been shown generically by \cite{david2022}. Hence, the eigenvalues of $H_{\rm dw}$ are either imaginary or consist of pairs of complex numbers with opposite real parts. In parameter space, the transition between propagating and non-propagative regions must then be an Exceptional Points at which the solution of two counter-propagating waves degenerates. 

\begin{figure}
    \centering
    \includegraphics[width=\columnwidth]{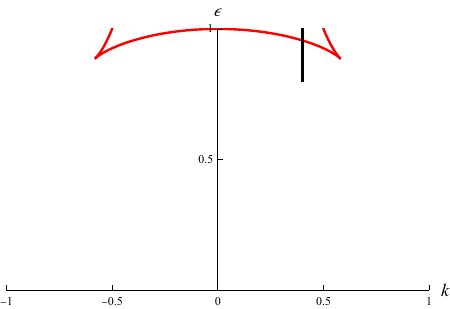}
    \includegraphics[width=\columnwidth]{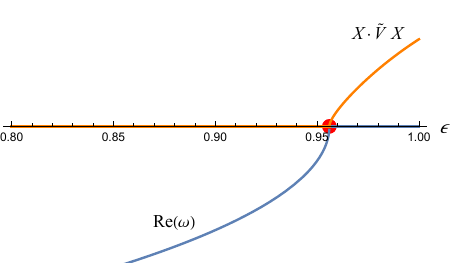}
    \caption{Same as Fig.~\ref{fig:toomreCollision} for the \textsc{Dustywave} problem. For ${k = 0.4}$, the Krein quantity $X\cdot \Tilde{V} X$ is represented as a function of $\epsilon$. \corr{This Krein quantity provides a quadratic constant of motion for dusty sound waves in the propagating phase (low $\epsilon$).}}
    \label{fig:dusty_phaseTransition}
\end{figure}

The characterization of these EPs is carried out according to the procedure described in \citet{delplace2021} (see App.~\ref{app:resultant} for details). Exceptional Points of order 2 (or EP2s) are located on a manifold of dimension 1 in the parameter space ($k,\epsilon$), which is a curve in a plane. Values of ($k,\epsilon$) for which an eigenvalue has a multiplicity of two are characterized by the fact that the characteristic polynomial $P(X)$ of $H_{\rm dw}$ has a root of multiplicity two. Such a root is therefore also a root of its derivative $P'(X)$, which means that the resultant $R_1 \equiv \mathcal{R}(P,P')$ between the two polynomials is 0. The resultant between two polynomials only cancels out if they have a common root. Since $H_{\rm dw}$ is anti-$\mathcal{PT}$-symmetric, $R_1$ is a real quantity. The values of ($k,\epsilon$), for which $R_1=0$, therefore define a curve. We calculate
\begin{equation}
    R_1 = -4k^6 (\epsilon-1)^2 \bigg(1 + k^4 - \epsilon + k^2 \big(2 + 9 \epsilon (-1 + \frac{3}{4} \epsilon)\big)\bigg),
    \label{eq:R1}
\end{equation}
and show on Fig.\ref{fig:dusty_phaseTransition} the curve $R_1=0$ (top panel, red). This curves provide a simple alternative derivation of the result of \cite{david2021}. Non-propagating waves exist for $\epsilon > 8/9$, which is indeed the minimum of the curve of EPs.


Interestingly, in addition to the EP2s, the above method reveals the existence of Exceptional Points of order 3 for which three eigenvectors merge. These EP3s form a fold of dimension 0, which are two points in ($k,\epsilon$) space. These two points are located at
\begin{equation}
    k = \pm \frac{1}{\sqrt{3}}, \quad \epsilon = \frac{8}{9},
\end{equation}
which are the points at which the curve of the EP2s loses its regularity. At these EP3s, $H_{\rm dw}$ has only two eigenvectors corresponding to the two eigenvalues 0 and $\frac{-i}{3}$. We note that these EP3s carry topological charges, also called winding numbers $W_3$,
\begin{equation}
    W_3\Big|_{\mathrm{EP3}_1} = +1,\quad
    W_3\Big|_{\mathrm{EP3}_2} = -1,
\end{equation}
where $\mathrm{EP3}_1 $ is the point $(k,\epsilon) = (+\frac{1}{\sqrt{3}},\frac{8}{9})$ and $\mathrm{EP3}_2 $ is the point $(k,\epsilon) = (-\frac{1}{\sqrt{3}},\frac{8}{9})$ (see App.~\ref{app:EP3s_charges}). The question of connecting these topological charges to edge modes or particular global modes remains open (e.g. \citealt{leclerc2023}).

According to the discussion in Sect. \ref{subsec:antiPT}, $H_{\rm dw}$ must be pseudo-chiral. The operator $\Tilde{V}$ can be found explicitly
\begin{eqnarray}
    \Tilde{V} &=& \left(
\begin{array}{cccc}
 \epsilon  & \epsilon -1 & i k & 0 \\
 \epsilon -1 & \frac{(\epsilon -1)^2}{\epsilon } & 0 & -\frac{i k (\epsilon -1)}{\epsilon } \\
 -i k & 0 & 0 & 0 \\
 0 & \frac{i k (\epsilon -1)}{\epsilon } & 0 & 0 \\
\end{array}
\right) . \label{eq:Uphbi}\\
\label{eq:dusty_pseudoChiral}
\end{eqnarray}
As has been shown, pseudo-chirality implies that if $\omega \neq -\omega^*$, the quantity $X\cdot \Tilde{V} X$ must be zero. This condition is fulfilled when sound propagates. Although the system is not conservative, it still possesses an associated Krein quantity $X\cdot \Tilde{V}  X$
\begin{eqnarray}
    X\cdot \Tilde{V}  X &=& \frac{1-\epsilon}{\epsilon}\vert\rho_\dust\vert^2 + \frac{\epsilon}{1-\epsilon} \vert\rho_\gas\vert^2 \\
    && - 2\left(\mathrm{Re}(\rho_\gas \rho_\dust)-k\frac{\rho_0}{c_\mathrm{s}}\mathrm{Im}(\rho_\gas v_\gas^* + \rho_\dust v_\dust^*)\right),\nonumber
    \label{eq:dusty_kreinSon}
\end{eqnarray}
where $\rho_0 = \rho_{\gas,0}+\rho_{\dust,0}$ is the total background density.
$X\cdot \Tilde{V}  X$ must be zero for propagating waves and non-zero for non-propagating waves. Figure~\ref{fig:dusty_phaseTransition} shows the numerical confirmation of this result. $X\cdot \Tilde{V}  X$ is only a constant of motion if it is zero, i.e. only for sound waves. Another technique, where the anti-$\mathcal{PT}$ operator is diagonalized, can be used to extract constants of motions in the unbroken phase, as it is done on the Kelvin-Helmholtz instability by \cite{qin2019} (see App. \ref{app:CP_operator} for details). \\
Finally, it should be noted that in a monofluid description of the mixture, the variables are $\rho = \rho_\gas+\rho_\dust$, $v = \rho_\gas v_\gas + \rho_\dust v_\dust$, $\frac{\rho_\dust}{\rho_\gas}$ and $\Delta v = v_\dust - v_\gas$ \citep{laibe2014dusty}.  If the perturbed quantities are denoted by $'$, the Krein quantity that cancels out for propagating perturbations is
\begin{equation}
\begin{split}
    & X\cdot\Tilde{V}_\mathrm{mono} X= \\ &
    \frac{(1-\epsilon)^3}{\epsilon}\left\vert\left(\frac{\rho_\dust}{\rho_\gas}\right)^\prime\right\vert^2+2k\mathrm{Im}\left((1-\epsilon)^2\left(\frac{\rho_\dust}{\rho_\gas}\right)^\prime \Delta v^{\prime *}+\frac{\rho^\prime}{\rho_0}v^{\prime *}\right).\nonumber    
\end{split}
\end{equation}
It is associated with the symmetry operator
\begin{equation}
    \Tilde{V}_\mathrm{mono} = \left(
\begin{array}{cccc}
 0 & 0 & -i k & 0 \\
 0 & \frac{(\epsilon -1)^3}{\epsilon } & 0 & -i k (\epsilon -1)^2 \\
 i k & 0 & 0 & 0 \\
 0 & i k (\epsilon -1)^2 & 0 & 0 \\
\end{array}
\right).
\label{eq:dustywave_sym}
\end{equation}

\corr{\cite{squire2018} have shown that when grains are additionally streaming though the gas, a generic resonant drag instability develops in the mixture. The system is neither $\mathcal{PT}$ nor anti-$\mathcal{PT}$ symmetric and Krein theory can not be applied directly. The instability comes from a more complex mechanism where a resonance occurs between an anti-$\mathcal{PT}$ symmetric correction due to drag to a leading $\mathcal{PT}$ symmetric perturbation that arises from the streaming (e.g. \citealt{zhuravlev2019,magnan2024}). Analysis of these problems from the point of view of discrete symmetries requires further specific studies.}

\section{Gravitational instability of a dusty disc}
\label{sec:discuss}

To our knowledge, there are still no theoretical predictions about what happens when systems with different discrete symmetries are combined. To illustrate this, we consider a final example, drawing on the analysis of \citet{longarini2022}, which deals with the gravitational instability of a dusty razor-thin disc. In the context of planet formation, the aim is to quantify the ways in which dust can favor the local collapse of gas or even clump itself into planetary embryos. This problem is a combination of the two problems presented in Sect. ~\ref{sec:toomre} and Sect. ~\ref{sec:DW_pb} and as such, is of order 6. With the notations used above, the matrix of linear perturbation in the local shearing box is  
\begin{align}
    &H_{\rm sgd} = \nonumber\\
    &\left(
\begin{array}{cccccc}
 0 & i k \Sigma _g & 0 & 0 & 0 & 0 \\
 -i k \left(\frac{2 \pi \mathcal{G}}{| k| }-\frac{c_g^2}{\Sigma _g}\right) & \frac{\epsilon }{t_s} & -2 \Omega  & -\frac{2 i \pi \mathcal{G} k}{| k| } & -\frac{\epsilon }{t_s} & 0 \\
 0 & -2 B & \frac{\epsilon }{t_s} & 0 & 0 & -\frac{\epsilon }{t_s} \\
 0 & 0 & 0 & 0 & i k \Sigma _d & 0 \\
 -\frac{2 i \pi \mathcal{G} k}{| k| } & -\frac{1}{t_s} & 0 & -i k \left(\frac{2 \pi \mathcal{G}}{| k| }-\frac{c_d^2}{\Sigma _d}\right) & \frac{1}{t_s} & -2 \Omega  \\
 0 & 0 & -\frac{1}{t_s} & 0 & -2 B & \frac{1}{t_s} \\
\end{array}
\right),
\end{align}
 where $\Omega$ is the orbital frequency and $B$ is the local Oort parameter. $c_{\rm g}$ and $c_{\rm d}$ denote the sound speeds of the gas and dust respectively. Time and lengths are rescaled to the stopping time $t_{\rm s}$ and the stopping length $c_{\rm g} t_{\rm s}$. The characteristic polynomial of the system is of order 5, so that the roots $\omega\left( k \right)$ cannot be determined analytically. In the regime of weak drag, \citet{longarini2022} determines the marginal stability numerically.

On the other hand, this problem is $\mathcal{PT}$-symmetric, with the operator ${U = \mathrm{diag}(-1,1,1,-1,1,1)}$. The marginal stability curve is therefore a curve of EPs that corresponds to the spontaneous breaking of $\mathcal{PT}$ symmetry. From the analysis above, its exact expression is
\begin{equation}
    \mathcal{R}_{\rm sgd}(P,P')=0,
    \label{eq:sgd}
\end{equation}
a polynomial of high order with respect to the parameters of the problem. \corr{The explicit expression of Eq.~\eqref{eq:sgd} extends over two pages and is given in the worksheet in the Acknowledgements. Hence, from a general perspective, the $\mathcal{PT}$-symmetric characterization of a marginal stability criterion allows to derive directly its analytical expression, albeit cumbersomely, without approximations relying on asymptotically strong or weak drag regimes.} Following \citet{qin2021}, this astrophysical example shows that the analysis of discrete symmetries therefore provides a powerful way to make analytical predictions about the marginal stability of symmetric systems with high dimensionalities.

\begin{figure}
    \centering
    \includegraphics[width=\columnwidth]{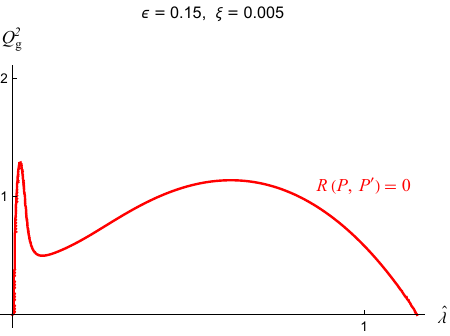}
    \caption{Exceptional Points of the dusty self-gravitating disk are a consequence of the spontaneous breaking of $\mathcal{PT}$ symmetry. The equation $R\left(P,P' \right) = 0$ is obtained analytically with a resultant method, reproducing the results of \cite{longarini2022}.}
    \label{fig:dustyToomre}
\end{figure}

\section{Conclusion}
For several astrophysical objects, the properties of the small perturbations around the equilibrium are controlled by discrete symmetries, such as the $\mathcal{PT}$ symmetry or the anti-$\mathcal{PT}$ symmetry. For example, we show that such symmetries are relevant for 
\begin{itemize}
\item The stability of isothermal spheres and thin self-gravitating disc: the Jeans and the are Toomre problem are bi-symmetric due to self-gravity.
\item The propagation of waves and the onset of convection in stratified fluids for compressible and adiabatic perturbations: the Schwarzschild problem is bi-symmetric due to buoyancy.
\item The propagation of waves and the onset of convection in stratified fluids for incompressible and diffusive perturbations: the Rayleigh-Bénard problem is anti-$\mathcal{PT}$-symmetric due to diffusion, as well as anti-Hermitian.
\item The propagation of a wave in a dust-gas mixture: the \textsc{dustywave} problem is anti-$\mathcal{PT}$ symmetric due to drag.
\end{itemize}
In these systems, the parameter space can be divided into different regions, in which the discrete symmetries are either broken or not. These regions are separated by Exceptional Points (EPs) for which the system of linear perturbation can no longer be diagonalised.

\corr{From a methodology perspective, the analysis of discrete symmetries allows when relevant to determine directly physical quantities that may be harder to obtain alternatively:
\begin{itemize}
\item The location of the EPs in the parameter space
can be determined analytically from the symmetries without knowing the dispersion relation. This method is particularly effective for systems with high dimensions, where the dispersion relations may be untractable for analytical techniques, as in the case of the dusty self-gravitating disc. 
\item Krein theory then makes it possible to directly determine the exchange of stability when EPs are crossed, as well as the partition of energy enforced by Krein invariants associated with the discrete symmetries, even for dissipative systems.
\end{itemize}
}
 Some of these EPs can be of high order and carry topological charges, as this is the case for the \textsc{dustywave} problem. Note that not all physical systems allow such discrete symmetries. In particular, their spontaneous breaking is not a necessary condition for the system to be linearly unstable, as the Rayleigh-Bénard problem shows. We have focused here on the analysis of local stability, which applies to homogeneous systems. The study of global modes for inhomogeneous systems requires an extension of the Wigner-Weyl framework to non-Hermitian systems (e.g. \citealt{onuki2020,perez2021,perez2022,perez2022thesis,leclerc2023}).

\section*{Acknowledgements}
We thank T. David--Cléris, E. Lynch, C. Longarini, P. Delplace, S. Labrosse and Y. Ricard for stimulating discussions and comments. We thank the conference \textsc{Star@Lyon} for stimulation of this study. GL, AL and NP acknowledge the support of the European Research Council (ERC) CoG project PODCAST No. 864965. AL is funded by Contrat Doctoral Spécifique Normalien. We have used \textsc{Mathematica} \citep{Mathematica}, worksheets can be found at \url{https://github.com/ArmandLeclerc/PTsymInAstroWaves}.

\bibliographystyle{plainnat}
\typeout{}
\bibliography{PTsym_in_astro}{}

\begin{thebibliography}{66}
\providecommand{\natexlab}[1]{#1}
\providecommand{\url}[1]{\texttt{#1}}
\expandafter\ifx\csname urlstyle\endcsname\relax
  \providecommand{\doi}[1]{doi: #1}\else
  \providecommand{\doi}{doi: \begingroup \urlstyle{rm}\Url}\fi

\bibitem[Ahuja(1973)]{ahuja1973}
Avtar~S Ahuja.
\newblock Wave equation and propagation parameters for sound propagation in
  suspensions.
\newblock \emph{Journal of Applied Physics}, 44\penalty0 (11):\penalty0
  4863--4868, 1973.

\bibitem[{Armitage}(2010)]{armitage2010}
Philip~J. {Armitage}.
\newblock \emph{{Astrophysics of Planet Formation}}.
\newblock Cambridge University Press, 2010.

\bibitem[Baines et~al.(1965)Baines, Williams, Asebiomo, and Agacy]{baines1965}
MJ~Baines, IP~Williams, AS~Asebiomo, and RL~Agacy.
\newblock Resistance to the motion of a small sphere moving through a gas.
\newblock \emph{Monthly Notices of the Royal Astronomical Society},
  130\penalty0 (1):\penalty0 63--74, 1965.

\bibitem[B{\'e}nard(1900)]{benard1900}
Henri B{\'e}nard.
\newblock Les tourbillons cellulaires dans une nappe liquide.
\newblock \emph{Revue Gen. Sci. Pure Appl.}, 11:\penalty0 1261--1271, 1900.

\bibitem[Bender and Boettcher(1998)]{bender1998}
Carl~M. Bender and Stefan Boettcher.
\newblock Real spectra in non-hermitian hamiltonians having $pt$ symmetry.
\newblock \emph{Phys. Rev. Lett.}, 80:\penalty0 5243--5246, Jun 1998.
\newblock \doi{10.1103/PhysRevLett.80.5243}.
\newblock URL \url{https://link.aps.org/doi/10.1103/PhysRevLett.80.5243}.

\bibitem[Bender et~al.(1999)Bender, Boettcher, and Meisinger]{bender1999}
Carl~M. Bender, Stefan Boettcher, and Peter~N. Meisinger.
\newblock {{PT-symmetric}} quantum mechanics.
\newblock \emph{Journal of Mathematical Physics}, 40\penalty0 (5):\penalty0
  2201--2229, May 1999.
\newblock ISSN 0022-2488.
\newblock \doi{10.1063/1.532860}.

\bibitem[Bender et~al.(2002)Bender, Brody, and Jones]{bender2002}
Carl~M. Bender, Dorje~C. Brody, and Hugh~F. Jones.
\newblock Complex {{Extension}} of {{Quantum Mechanics}}.
\newblock \emph{Physical Review Letters}, 89\penalty0 (27):\penalty0 270401,
  December 2002.
\newblock \doi{10.1103/PhysRevLett.89.270401}.

\bibitem[Bender et~al.(2004)Bender, Brody, and Jones]{bender2004}
Carl~M. Bender, Dorje~C. Brody, and Hugh~F. Jones.
\newblock Extension of \$\textbackslash
  mathcal\{\vphantom\}{{PT}}\vphantom\{\}\$-symmetric quantum mechanics to
  quantum field theory with cubic interaction.
\newblock \emph{Physical Review D}, 70\penalty0 (2):\penalty0 025001, July
  2004.
\newblock \doi{10.1103/PhysRevD.70.025001}.

\bibitem[Berg{\'e} and Dubois(1984)]{berge1984}
P~Berg{\'e} and M~Dubois.
\newblock Rayleigh-b{\'e}nard convection.
\newblock \emph{Contemporary Physics}, 25\penalty0 (6):\penalty0 535--582,
  1984.

\bibitem[{Bertin} and {Lodato}(1999)]{bertin1999}
G.~{Bertin} and G.~{Lodato}.
\newblock {A class of self-gravitating accretion disks}.
\newblock \emph{\aap}, 350:\penalty0 694--704, October 1999.
\newblock \doi{10.48550/arXiv.astro-ph/9908095}.

\bibitem[Bodenschatz et~al.(2000)Bodenschatz, Pesch, and
  Ahlers]{bodenschatz2000}
Eberhard Bodenschatz, Werner Pesch, and Guenter Ahlers.
\newblock Recent developments in rayleigh-b{\'e}nard convection.
\newblock \emph{Annual review of fluid mechanics}, 32\penalty0 (1):\penalty0
  709--778, 2000.

\bibitem[Brandenburg(2021)]{brandenburg2021}
Axel Brandenburg.
\newblock Lecture 4: Rayleigh–bénard problem.
\newblock
  {\url{http://norlx51.nordita.org/~brandenb/teach/AdvAstroFluids/4_Convection/notes.pdf}},
  2021.

\bibitem[Cowling(1941)]{cowling1941}
Thomas~G Cowling.
\newblock The non-radial oscillations of polytropic stars.
\newblock \emph{Monthly Notices of the Royal Astronomical Society, Vol. 101, p.
  367}, 101:\penalty0 367, 1941.

\bibitem[David et~al.(2022)David, Delplace, and Venaille]{david2022}
Tomos~W. David, Pierre Delplace, and Antoine Venaille.
\newblock How do discrete symmetries shape the stability of geophysical flows?
\newblock \emph{Physics of Fluids}, 34\penalty0 (5):\penalty0 056605, May 2022.
\newblock ISSN 1070-6631.
\newblock \doi{10.1063/5.0088936}.

\bibitem[David-Cl{\'e}ris and Laibe(2021)]{david2021}
Timoth{\'e}e David-Cl{\'e}ris and Guillaume Laibe.
\newblock Large dust fractions can prevent the propagation of soundwaves.
\newblock \emph{Monthly Notices of the Royal Astronomical Society},
  504\penalty0 (2):\penalty0 2889--2894, 2021.

\bibitem[Delplace et~al.(2021)Delplace, Yoshida, and Hatsugai]{delplace2021}
Pierre Delplace, Tsuneya Yoshida, and Yasuhiro Hatsugai.
\newblock Symmetry-protected multifold exceptional points and their topological
  characterization.
\newblock \emph{Physical Review Letters}, 127\penalty0 (18):\penalty0 186602,
  2021.

\bibitem[{El-Ganainy} et~al.(2007){El-Ganainy}, Makris, Christodoulides, and
  Musslimani]{el-ganainy2007}
R.~{El-Ganainy}, K.~G. Makris, D.~N. Christodoulides, and Ziad~H. Musslimani.
\newblock Theory of coupled optical {{PT-symmetric}} structures.
\newblock \emph{Optics Letters}, 32\penalty0 (17):\penalty0 2632--2634,
  September 2007.
\newblock ISSN 1539-4794.
\newblock \doi{10.1364/OL.32.002632}.

\bibitem[{El-Ganainy} et~al.(2018){El-Ganainy}, Makris, Khajavikhan,
  Musslimani, Rotter, and Christodoulides]{el-ganainy2018}
Ramy {El-Ganainy}, Konstantinos~G. Makris, Mercedeh Khajavikhan, Ziad~H.
  Musslimani, Stefan Rotter, and Demetrios~N. Christodoulides.
\newblock Non-{{Hermitian}} physics and {{PT}} symmetry.
\newblock \emph{Nature Physics}, 14\penalty0 (1):\penalty0 11--19, January
  2018.
\newblock ISSN 1745-2481.
\newblock \doi{10.1038/nphys4323}.

\bibitem[Feng et~al.(2013)Feng, Xu, Fegadolli, Lu, Oliveira, Almeida, Chen, and
  Scherer]{feng2013}
Liang Feng, Ye-Long Xu, William~S. Fegadolli, Ming-Hui Lu, Jos{\'e} E.~B.
  Oliveira, Vilson~R. Almeida, Yan-Feng Chen, and Axel Scherer.
\newblock Experimental demonstration of a unidirectional reflectionless
  parity-time metamaterial at optical frequencies.
\newblock \emph{Nature Materials}, 12\penalty0 (2):\penalty0 108--113, February
  2013.
\newblock ISSN 1476-4660.
\newblock \doi{10.1038/nmat3495}.

\bibitem[Feng et~al.(2014)Feng, Wong, Ma, Wang, and Zhang]{feng2014}
Liang Feng, Zi~Jing Wong, Ren-Min Ma, Yuan Wang, and Xiang Zhang.
\newblock Single-mode laser by parity-time symmetry breaking.
\newblock \emph{Science}, 346\penalty0 (6212):\penalty0 972--975, November
  2014.
\newblock \doi{10.1126/science.1258479}.

\bibitem[Feng et~al.(2017)Feng, {El-Ganainy}, and Ge]{feng2017}
Liang Feng, Ramy {El-Ganainy}, and Li~Ge.
\newblock Non-{{Hermitian}} photonics based on parity\textendash time symmetry.
\newblock \emph{Nature Photonics}, 11\penalty0 (12):\penalty0 752--762,
  December 2017.
\newblock ISSN 1749-4893.
\newblock \doi{10.1038/s41566-017-0031-1}.

\bibitem[Fu and Qin(2020{\natexlab{a}})]{fu2020}
Yichen Fu and Hong Qin.
\newblock The physics of spontaneous parity-time symmetry breaking in the
  kelvin--helmholtz instability.
\newblock \emph{New Journal of Physics}, 22\penalty0 (8):\penalty0 083040,
  2020{\natexlab{a}}.

\bibitem[Fu and Qin(2020{\natexlab{b}})]{fu2020d}
Yichen Fu and Hong Qin.
\newblock The physics of spontaneous parity-time symmetry breaking in the
  {{Kelvin}}\textendash{{Helmholtz}} instability.
\newblock \emph{New Journal of Physics}, 22\penalty0 (8):\penalty0 083040,
  August 2020{\natexlab{b}}.
\newblock ISSN 1367-2630.
\newblock \doi{10.1088/1367-2630/aba38f}.

\bibitem[Fu and Qin(2023)]{fu2023}
Yichen Fu and Hong Qin.
\newblock Topological modes and spectral flows in inhomogeneous
  {{PT-symmetric}} continuous media, September 2023.

\bibitem[Ge and T{\"u}reci(2013)]{ge2013}
Li~Ge and Hakan~E T{\"u}reci.
\newblock Antisymmetric pt-photonic structures with balanced positive-and
  negative-index materials.
\newblock \emph{Physical Review A}, 88\penalty0 (5):\penalty0 053810, 2013.

\bibitem[Goodman(2003)]{goodman2003}
Jeremy Goodman.
\newblock Self-gravity and quasi-stellar object discs.
\newblock \emph{Monthly Notices of the Royal Astronomical Society},
  339\penalty0 (4):\penalty0 937--948, 2003.

\bibitem[Goodman and Narayan(1988)]{goodman1988}
Jeremy Goodman and Ramesh Narayan.
\newblock The stability of accretion tori--iii. the effect of self-gravity.
\newblock \emph{Monthly Notices of the Royal Astronomical Society},
  231\penalty0 (1):\penalty0 97--114, 1988.

\bibitem[Gumerov et~al.(1988)Gumerov, Ivandaev, and Nigmatulin]{gumerov1988}
NA~Gumerov, AI~Ivandaev, and RI~Nigmatulin.
\newblock Sound waves in monodisperse gas-particle or vapour-droplet mixtures.
\newblock \emph{Journal of Fluid Mechanics}, 193:\penalty0 53--74, 1988.

\bibitem[Guo et~al.(2009)Guo, Salamo, Duchesne, Morandotti, {Volatier-Ravat},
  Aimez, Siviloglou, and Christodoulides]{guo2009}
A.~Guo, G.~J. Salamo, D.~Duchesne, R.~Morandotti, M.~{Volatier-Ravat},
  V.~Aimez, G.~A. Siviloglou, and D.~N. Christodoulides.
\newblock Observation of \$\textbackslash
  mathcal\{\vphantom\}{{P}}\vphantom\{\}\textbackslash
  mathcal\{\vphantom\}{{T}}\vphantom\{\}\$-{{Symmetry Breaking}} in {{Complex
  Optical Potentials}}.
\newblock \emph{Physical Review Letters}, 103\penalty0 (9):\penalty0 093902,
  August 2009.
\newblock \doi{10.1103/PhysRevLett.103.093902}.

\bibitem[Hodaei et~al.(2017)Hodaei, Hassan, Wittek, Garcia-Gracia, El-Ganainy,
  Christodoulides, and Khajavikhan]{hodaei2017}
Hossein Hodaei, Absar~U Hassan, Steffen Wittek, Hipolito Garcia-Gracia, Ramy
  El-Ganainy, Demetrios~N Christodoulides, and Mercedeh Khajavikhan.
\newblock Enhanced sensitivity at higher-order exceptional points.
\newblock \emph{Nature}, 548\penalty0 (7666):\penalty0 187--191, 2017.

\bibitem[Jeans(1902)]{jeans1902}
James~Hopwood Jeans.
\newblock I. the stability of a spherical nebula.
\newblock \emph{Philosophical Transactions of the Royal Society of London.
  Series A, Containing Papers of a Mathematical or Physical Character},
  199\penalty0 (312-320):\penalty0 1--53, 1902.

\bibitem[Kirillov(2013)]{kirillov2013}
Oleg~N Kirillov.
\newblock Nonconservative stability problems of modern physics.
\newblock In \emph{Nonconservative Stability Problems of Modern Physics}. de
  Gruyter, 2013.

\bibitem[Klaiman et~al.(2008)Klaiman, G{\"u}nther, and Moiseyev]{klaiman2008}
Shachar Klaiman, Uwe G{\"u}nther, and Nimrod Moiseyev.
\newblock Visualization of {{Branch Points}} in \$\textbackslash
  mathcal\{\vphantom\}{{P}}\vphantom\{\}\textbackslash
  mathcal\{\vphantom\}{{T}}\vphantom\{\}\$-{{Symmetric Waveguides}}.
\newblock \emph{Physical Review Letters}, 101\penalty0 (8):\penalty0 080402,
  August 2008.
\newblock \doi{10.1103/PhysRevLett.101.080402}.

\bibitem[Kre{\u\i}n(1950)]{krein1950}
MG~Kre{\u\i}n.
\newblock A generalization of some investigations on linear differential
  equations with periodic coefficients.
\newblock In \emph{Dokl. Akad. Nauk SSSR A}, volume~73, page 445, 1950.

\bibitem[Laibe and Price(2011)]{laibe2011}
Guillaume Laibe and Daniel~J. Price.
\newblock {dustybox and dustywave: two test problems for numerical simulations
  of two-fluid astrophysical dustâ€“gas mixtures}.
\newblock \emph{Monthly Notices of the Royal Astronomical Society},
  418\penalty0 (3):\penalty0 1491--1497, 12 2011.
\newblock ISSN 0035-8711.
\newblock \doi{10.1111/j.1365-2966.2011.19291.x}.
\newblock URL \url{https://doi.org/10.1111/j.1365-2966.2011.19291.x}.

\bibitem[Laibe and Price(2014)]{laibe2014dusty}
Guillaume Laibe and Daniel~J Price.
\newblock Dusty gas with one fluid.
\newblock \emph{Monthly Notices of the Royal Astronomical Society},
  440\penalty0 (3):\penalty0 2136--2146, 2014.

\bibitem[Leclerc et~al.(2023)Leclerc, Jezequel, Perez, Bhandare, Laibe, and
  Delplace]{leclerc2023}
Armand Leclerc, Lucien Jezequel, Nicolas Perez, Asmita Bhandare, Guillaume
  Laibe, and Pierre Delplace.
\newblock The exceptional ring of buoyancy instability in stars, 2023.

\bibitem[Lin et~al.(2011)Lin, Ramezani, Eichelkraut, Kottos, Cao, and
  Christodoulides]{lin2011}
Zin Lin, Hamidreza Ramezani, Toni Eichelkraut, Tsampikos Kottos, Hui Cao, and
  Demetrios~N. Christodoulides.
\newblock Unidirectional {{Invisibility Induced}} by \$\textbackslash
  mathcal\{\vphantom\}{{P}}\vphantom\{\}\textbackslash
  mathcal\{\vphantom\}{{T}}\vphantom\{\}\$-{{Symmetric Periodic Structures}}.
\newblock \emph{Physical Review Letters}, 106\penalty0 (21):\penalty0 213901,
  May 2011.
\newblock \doi{10.1103/PhysRevLett.106.213901}.

\bibitem[Longarini et~al.(2022)Longarini, Lodato, Bertin, and
  Armitage]{longarini2022}
Cristiano Longarini, Giuseppe Lodato, Giuseppe Bertin, and Philip~J Armitage.
\newblock The role of the drag force in the gravitational stability of dusty
  planet forming disc {\textendash} i. analytical theory.
\newblock \emph{Monthly Notices of the Royal Astronomical Society},
  519\penalty0 (2):\penalty0 2017--2029, dec 2022.
\newblock \doi{10.1093/mnras/stac3653}.
\newblock URL \url{https://doi.org/10.1093%2Fmnras%2Fstac3653}.

\bibitem[Magnan et~al.(2024)Magnan, Heinemann, and Latter]{magnan2024}
Nathan Magnan, Tobias Heinemann, and Henrik~N Latter.
\newblock A physical picture for the acoustic resonant drag instability.
\newblock \emph{Monthly Notices of the Royal Astronomical Society}, page
  stae052, 2024.

\bibitem[Makris et~al.(2008)Makris, {El-Ganainy}, Christodoulides, and
  Musslimani]{makris2008}
K.~G. Makris, R.~{El-Ganainy}, D.~N. Christodoulides, and Z.~H. Musslimani.
\newblock Beam {{Dynamics}} in \$\textbackslash
  mathcal\{\vphantom\}{{P}}\vphantom\{\}\textbackslash
  mathcal\{\vphantom\}{{T}}\vphantom\{\}\$ {{Symmetric Optical Lattices}}.
\newblock \emph{Physical Review Letters}, 100\penalty0 (10):\penalty0 103904,
  March 2008.
\newblock \doi{10.1103/PhysRevLett.100.103904}.

\bibitem[Mostafazadeh(2002{\natexlab{a}})]{mostafazadeh2002a}
Ali Mostafazadeh.
\newblock Pseudo-hermiticity versus pt symmetry: the necessary condition for
  the reality of the spectrum of a non-hermitian hamiltonian.
\newblock \emph{Journal of Mathematical Physics}, 43\penalty0 (1):\penalty0
  205--214, 2002{\natexlab{a}}.

\bibitem[Mostafazadeh(2002{\natexlab{b}})]{mostafazadeh2002b}
Ali Mostafazadeh.
\newblock Pseudo-hermiticity versus pt-symmetry iii: Equivalence of
  pseudo-hermiticity and the presence of antilinear symmetries.
\newblock \emph{Journal of Mathematical Physics}, 43\penalty0 (8):\penalty0
  3944--3951, 2002{\natexlab{b}}.

\bibitem[Mostafazadeh and Batal(2004)]{mostafazadeh2004}
Ali Mostafazadeh and Ahmet Batal.
\newblock Physical aspects of pseudo-{{Hermitian}} and {{PT-symmetric}} quantum
  mechanics.
\newblock \emph{Journal of Physics A: Mathematical and General}, 37\penalty0
  (48):\penalty0 11645, November 2004.
\newblock ISSN 0305-4470.
\newblock \doi{10.1088/0305-4470/37/48/009}.

\bibitem[Musslimani et~al.(2008)Musslimani, Makris, El-Ganainy, and
  Christodoulides]{musslimani2008}
Ziad~H Musslimani, Konstantinos~G Makris, Ramy El-Ganainy, and Demetrios~N
  Christodoulides.
\newblock Optical solitons in p t periodic potentials.
\newblock \emph{Physical Review Letters}, 100\penalty0 (3):\penalty0 030402,
  2008.

\bibitem[Onuki(2020)]{onuki2020}
Yohei Onuki.
\newblock Quasi-local method of wave decomposition in a slowly varying medium.
\newblock \emph{Journal of Fluid Mechanics}, 883:\penalty0 A56, 2020.

\bibitem[{\"O}zdemir et~al.(2019){\"O}zdemir, Rotter, Nori, and
  Yang]{ozdemir2019}
{\c{S}}ahin~Kaya {\"O}zdemir, Stefan Rotter, Franco Nori, and L~Yang.
\newblock Parity--time symmetry and exceptional points in photonics.
\newblock \emph{Nature materials}, 18\penalty0 (8):\penalty0 783--798, 2019.

\bibitem[Peng et~al.(2014)Peng, {\"O}zdemir, Lei, Monifi, Gianfreda, Long, Fan,
  Nori, Bender, and Yang]{peng2014}
Bo~Peng, {\c S}ahin~Kaya {\"O}zdemir, Fuchuan Lei, Faraz Monifi, Mariagiovanna
  Gianfreda, Gui~Lu Long, Shanhui Fan, Franco Nori, Carl~M. Bender, and Lan
  Yang.
\newblock Parity\textendash time-symmetric whispering-gallery microcavities.
\newblock \emph{Nature Physics}, 10\penalty0 (5):\penalty0 394--398, May 2014.
\newblock ISSN 1745-2481.
\newblock \doi{10.1038/nphys2927}.

\bibitem[Perez(2022)]{perez2022thesis}
Nicolas Perez.
\newblock \emph{Topological waves in geophysical and astrophysical fluids}.
\newblock PhD thesis, Ecole normale sup{\'e}rieure de lyon-ENS LYON, 2022.

\bibitem[Perez et~al.(2021)Perez, Delplace, and Venaille]{perez2021}
Nicolas Perez, Pierre Delplace, and Antoine Venaille.
\newblock Manifestation of the berry curvature in geophysical ray tracing.
\newblock \emph{Proceedings of the Royal Society A}, 477\penalty0
  (2248):\penalty0 20200844, 2021.

\bibitem[Perez et~al.(2022)Perez, Delplace, and Venaille]{perez2022}
Nicolas Perez, Pierre Delplace, and Antoine Venaille.
\newblock Unidirectional modes induced by nontraditional coriolis force in
  stratified fluids.
\newblock \emph{Physical Review Letters}, 128\penalty0 (18):\penalty0 184501,
  2022.

\bibitem[Pringle and King(2007)]{pringle2007}
James~E Pringle and Andrew King.
\newblock \emph{Astrophysical flows}.
\newblock Cambridge University Press, 2007.

\bibitem[Qin et~al.(2019)Qin, Zhang, Glasser, and Xiao]{qin2019}
Hong Qin, Ruili Zhang, Alexander~S. Glasser, and Jianyuan Xiao.
\newblock Kelvin-{{Helmholtz}} instability is the result of parity-time
  symmetry breaking.
\newblock \emph{Physics of Plasmas}, 26\penalty0 (3):\penalty0 032102, March
  2019.
\newblock ISSN 1070-664X.
\newblock \doi{10.1063/1.5088498}.

\bibitem[Qin et~al.(2021)Qin, Fu, Glasser, and Yahalom]{qin2021}
Hong Qin, Yichen Fu, Alexander~S. Glasser, and Asher Yahalom.
\newblock Spontaneous and explicit parity-time-symmetry breaking in drift-wave
  instabilities.
\newblock \emph{Physical Review E}, 104\penalty0 (1):\penalty0 015215, July
  2021.
\newblock \doi{10.1103/PhysRevE.104.015215}.

\bibitem[Rayleigh(1916)]{rayleigh1916}
Lord Rayleigh.
\newblock Lix. on convection currents in a horizontal layer of fluid, when the
  higher temperature is on the under side.
\newblock \emph{The London, Edinburgh, and Dublin Philosophical Magazine and
  Journal of Science}, 32\penalty0 (192):\penalty0 529--546, 1916.

\bibitem[Regensburger et~al.(2012)Regensburger, Bersch, Miri, Onishchukov,
  Christodoulides, and Peschel]{regensburger2012}
Alois Regensburger, Christoph Bersch, Mohammad-Ali Miri, Georgy Onishchukov,
  Demetrios~N. Christodoulides, and Ulf Peschel.
\newblock Parity\textendash time synthetic photonic lattices.
\newblock \emph{Nature}, 488\penalty0 (7410):\penalty0 167--171, August 2012.
\newblock ISSN 1476-4687.
\newblock \doi{10.1038/nature11298}.

\bibitem[R{\"u}ter et~al.(2010)R{\"u}ter, Makris, {El-Ganainy},
  Christodoulides, Segev, and Kip]{ruter2010}
Christian~E. R{\"u}ter, Konstantinos~G. Makris, Ramy {El-Ganainy}, Demetrios~N.
  Christodoulides, Mordechai Segev, and Detlef Kip.
\newblock Observation of parity\textendash time symmetry in optics.
\newblock \emph{Nature Physics}, 6\penalty0 (3):\penalty0 192--195, March 2010.
\newblock ISSN 1745-2481.
\newblock \doi{10.1038/nphys1515}.

\bibitem[Saffman(1962)]{saffman1962}
PG~Saffman.
\newblock On the stability of laminar flow of a dusty gas.
\newblock \emph{Journal of fluid mechanics}, 13\penalty0 (1):\penalty0
  120--128, 1962.

\bibitem[{Schwarzschild}(1906)]{schwarzschild1906}
K.~{Schwarzschild}.
\newblock {On the equilibrium of the Sun's atmosphere}.
\newblock \emph{Nachrichten von der K{\"o}niglichen Gesellschaft der
  Wissenschaften zu G{\"o}ttingen. Math.-phys. Klasse}, 195:\penalty0 41--53,
  January 1906.

\bibitem[Squire and Hopkins(2018)]{squire2018}
Jonathan Squire and Philip~F Hopkins.
\newblock Resonant drag instability of grains streaming in fluids.
\newblock \emph{The Astrophysical Journal Letters}, 856\penalty0 (1):\penalty0
  L15, 2018.

\bibitem[Toomre(1964)]{toomre1964}
Alar Toomre.
\newblock On the gravitational stability of a disk of stars.
\newblock \emph{Astrophysical Journal, vol. 139, p. 1217-1238 (1964).},
  139:\penalty0 1217--1238, 1964.

\bibitem[{Wolfram Research Inc.}(2021)]{Mathematica}
{Wolfram Research Inc.}
\newblock Mathematica, {V}ersion 13.0.0, 2021.
\newblock URL \url{https://www.wolfram.com/mathematica}.
\newblock Champaign, IL.

\bibitem[Zhang et~al.(2020{\natexlab{a}})Zhang, Feng, Chen, Ge, and
  Wan]{zhang2020synthetic}
Fangxing Zhang, Yaming Feng, Xianfeng Chen, Li~Ge, and Wenjie Wan.
\newblock Synthetic anti-pt symmetry in a single microcavity.
\newblock \emph{Physical review letters}, 124\penalty0 (5):\penalty0 053901,
  2020{\natexlab{a}}.

\bibitem[Zhang et~al.(2020{\natexlab{b}})Zhang, Huang, Zhang, Li, Qiu, Nori,
  and Jing]{zhang2020breaking}
Huilai Zhang, Ran Huang, Sheng-Dian Zhang, Ying Li, Cheng-Wei Qiu, Franco Nori,
  and Hui Jing.
\newblock Breaking anti-pt symmetry by spinning a resonator.
\newblock \emph{Nano Letters}, 20\penalty0 (10):\penalty0 7594--7599,
  2020{\natexlab{b}}.

\bibitem[Zhang et~al.(2020{\natexlab{c}})Zhang, Qin, and Xiao]{zhang2020pt}
Ruili Zhang, Hong Qin, and Jianyuan Xiao.
\newblock Pt-symmetry entails pseudo-hermiticity regardless of
  diagonalizability.
\newblock \emph{Journal of Mathematical Physics}, 61\penalty0 (1):\penalty0
  012101, 2020{\natexlab{c}}.

\bibitem[Zhuravlev(2019)]{zhuravlev2019}
VV~Zhuravlev.
\newblock On the nature of the resonant drag instability of dust streaming in
  protoplanetary disc.
\newblock \emph{Monthly Notices of the Royal Astronomical Society},
  489\penalty0 (3):\penalty0 3850--3869, 2019.

\end{thebibliography}

\appendix

\section{Rayleigh-Bénard: linear perturbations}
\label{app:rayleigh-benard}
Consider a two-dimensional fluid parametrised by the coordinates $\left(x,z \right)$. The fluid is at rest in a constant  gravity field $\mathbf{g} = - g \mathbf{u}_{z}$. It has constant density $\rho_0$ and the temperature profile has a constant gradient $T(z) = T_\mathrm{h}-\beta z$ with $\beta = (T_\mathrm{h}-T_\mathrm{c})/L$. The coefficient of thermal expansion of the fluid is $\alpha = -\rho_0^{-1}\left(\partial \rho / \partial T\right)_P$. The dynamical viscosity and thermal diffusivity of the fluid are denoted by $\nu$ and $\kappa$ respectively. The linear system describing the evolution of small perturbations in the Boussinesq approximation around this background state are
\begin{eqnarray}
    \partial_x v_x + \partial_z v_z &=& 0,\\
    (\partial_t - \nu \Delta)v_x &=& -\frac{1}{\rho_0}\partial_x p\\
    (\partial_t - \nu \Delta)v_z &=& -\frac{1}{\rho_0}\partial_z p + \alpha g T,\\
    (\partial_t - \kappa \Delta)T &=& \beta v_z .
\end{eqnarray}
Applying double curl, and enforcing the incompressibility equation $\nabla \cdot \mathbf{v} = 0$, the equation for $u_x$ decouples from the other two equations, leaving  
\begin{eqnarray}
    (\partial_t - \nu \Delta)\Delta v_z &=& \alpha g \partial_{xx}T,  \label{eq:rb_op1}\\
    (\partial_t - \kappa \Delta) T &=& \beta v_z .  \label{eq:rb_op2}
\end{eqnarray}
For plane wave solutions of the form $\exp{(-i\omega t + ik_z z + ik_x x)}$, Eqs.~\ref{eq:rb_op1} -- \ref{eq:rb_op2} become
\begin{eqnarray}
    \left(-i\omega - \nu (k_x^2+k_z^2)\right)(k_x^2+k_z^2) v_z &=& - \alpha g k_x^2 T,\\
    \left(-i\omega - \kappa (k_x^2+k_z^2)\right) T &=& \beta v_z .  
\end{eqnarray}
Rescaling the time by $L^2/\nu$, the lengths  by $L$ and the temperature with $\beta L$ leads to the introduction of the Rayleigh and Prandtl numbers, $\mathrm{Ra} = g\alpha\beta L^4 /\nu\kappa$ and $\mathrm{Pr} = \nu/\kappa$. The following rescaling of the equations $T \mapsto \frac{k_x}{k}\left(\frac{\mathrm{Ra}}{\mathrm{Pr}}\right)^{1/2} T$ with $k = \sqrt{k_x^2+k_z^2}$ provides a system of eigenvalue that has the symetric form
\begin{eqnarray}
    \omega X &=& H_{\rm RB}X,\\
    H_{\rm RB} &=& -i\begin{pmatrix}
        k^2 & -\frac{k_x}{k}\left(\frac{\mathrm{Ra}}{\mathrm{Pr}}\right)^{1/2}\\
        -\frac{k_x}{k}\left(\frac{\mathrm{Ra}}{\mathrm{Pr}}\right)^{1/2} & k^2/\mathrm{Pr}
    \end{pmatrix},
\end{eqnarray}
which corresponds to Eq.\ref{eq:rayleigh-benard-operator}. Consequently, $iH_{\rm RB}$ is Hermitian if and only if $\mathrm{Ra}\ge0$.

\section{\textsc{dustywave}: linear perturbations}
\label{app:dustyEq}
Consider a dusty mixture where gas and dust are treated as two fluid interacting with each other via a drag force. For pressureless dust in 1D, the equations for mass and momentum conservation are
\begin{eqnarray}
    (\partial_t + {v}_\gas \partial_x) {\rho}_\gas &=& -{\rho}_\gas\partial_x {v}_\gas,\\
    (\partial_t + {v}_\dust \partial_x) {\rho}_\dust &=& -{\rho}_\dust\partial_x {v}_\dust,\\
    (\partial_t + {v}_\gas \partial_x) {v}_\gas &=&  \frac{K}{{\rho}_\gas} ({v}_\dust-{v}_\gas)-\frac{1}{{\rho}_\gas}\partial_x {P} ,\\
    (\partial_t + {v}_\dust \partial_x) {v}_\dust &=&  \frac{K}{{\rho}_\dust} ({v}_\gas-{v}_\dust),
\end{eqnarray}
where $\rho_{\rm g}$, $\rho_{\rm d}$ denote the gas and the dust densities, $v_{\rm g}$, $v_{\rm d}$ denote the gas and the dust velocities, and $K$ is the drag coefficient. The mixture is initially homogeneous and at rest. Writing each quantity as a background and perturbation component $f = f_0 + f^\prime$. The system of equations can be put in a symmetric form by using the rescaled quantities $(\rho^\prime_\gas,\rho^\prime_\dust,v^\prime_\gas, v^\prime_\dust) \mapsto X = \begin{pmatrix}  \frac{1}{\rho_{\gas,0}+\rho_{\dust,0}}\rho^\prime_\gas ,& \frac{1}{\rho_{\gas,0}+\rho_{\dust,0}}\rho^\prime_\dust ,& \frac{\rho_{\gas,0}}{c_\mathrm{s}(\rho_{\gas,0}+\rho_{\dust,0})}v^\prime_\gas ,& \frac{\rho_{\dust,0}}{c_\mathrm{s}(\rho_{\gas,0}+\rho_{\dust,0})}v^\prime_\dust \end{pmatrix}^\top$. For isothermal pertubations ${P^\prime(x,t) = c_\mathrm{s}^2 \rho^\prime_\gas(x,t)}$, the evolution of the perturbations are given by the system
\begin{equation}
    \partial_t X = 
    \begin{pmatrix}
        0 & 0 & -c_\mathrm{s} \partial_x & 0 \\
        0 & 0 & 0 & -c_\mathrm{s}  \partial_x \\
        -c_\mathrm{s} \partial_x & 0 & -\frac{K}{\rho_{\gas,0}} & +\frac{(1-\epsilon)K}{\epsilon\rho_{\gas,0}} \\
        0 & 0 & \frac{\epsilon K}{(1-\epsilon)\rho_{\dust,0}} & -\frac{K}{\rho_{\dust,0}} 
    \end{pmatrix}X,
\end{equation}
where $\epsilon \equiv \frac{\rho_{\dust,0}}{\rho_{\gas,0}+\rho_{\dust,0}}$ is the dust fraction, and $1-\epsilon =  \frac{\rho_{\gas,0}}{\rho_{\gas,0}+\rho_{\dust,0}}$ is the gas fraction. The barycentric stopping time $t_\mathrm{s} \equiv \frac{\rho_{\gas,0}\rho_{\dust,0}}{K(\rho_{\gas,0}+\rho_{\dust,0})}$, corresponds to the time needed for drag to damp the differential velocity between gas and dust. The stopping length $c_{\rm s} t_{\rm s}$ is the typical spatial separation between the fluids. Using $t_\mathrm{s}$ and $c_\mathrm{s}t_\mathrm{s}$ as units of time and length, we obtain the following dimensionless system of equations
\begin{eqnarray} 
    H_{\rm dw} X &=& i\partial_t X,\\
    H_{\rm dw} &\equiv& \begin{pmatrix}
    0 & \;\;\; 0 &\;\;\;\; k & 0 \\
    0 &\;\;\; 0 &\;\;\;\; 0 & k \\
    k &\;\;\; 0 &\;\;\;\; i\epsilon  &  \;-i(1-\epsilon) \\
    0 &\;\;\; 0 &\; -i\epsilon  & \;\;\;i(1-\epsilon)
    \end{pmatrix} .
    \label{eqApp:eigen_dusty}
\end{eqnarray}

\section{ EP{\scriptsize s} from resultants}
\label{app:resultant}
\corr{
Consider two polynomials $Q1\left( X\right)$ and $Q2\left( X\right)$ defined by
\begin{eqnarray}
    Q_1\left( X\right) &=& {\displaystyle \sum_{i=1}^{n}} a_i X^i = a_n {\displaystyle \prod_{i=1}^{n}}(X-\alpha_i),\\
    Q_2\left( X\right) &=& {\displaystyle \sum_{j=1}^{m}} b_j X^j = b_m {\displaystyle \prod_{j=1}^{m}}(X-\beta_j),
\end{eqnarray}
where $\alpha_i$ and $\beta_j$ are their roots. The resultant $\mathcal{R}(Q_1,Q_2)$ of these two polynomials is defined as
\begin{equation}
\mathcal{R}(Q_1,Q_2) \equiv \det S(Q_1,Q_2),
\label{eq:defResultant}
\end{equation}
where $S(Q_1,Q_2)$ is the Sylvester matrix of $Q_1$ and $Q_2$. It satisfies
\begin{equation}
\mathcal{R}(Q_1,Q_2) = a_n^m b_m^n {\displaystyle \prod_{i,j}^{}}(\alpha_i-\beta_j),
\end{equation}
which cancels if and only if $Q_1$ and $Q_2$ share a common root. The quantity \eqref{eq:defResultant} is determined by the coefficients $a_i$, $b_j$ and does not require to find the roots $\alpha_i$, $\beta_j$.} Define $P'$, the derivative of $P(X)$. The cancellation of the resultant $R_1$ of $P$ and $P'$ characterize the fact that $P$ has a root with double multiplicity. In the context of this study, Exceptional Points of order at least 2 is the space formed by parameters for which 
\begin{equation}
    R_1 \equiv \mathcal{R}(P,P') = 0.
\end{equation}

\section{\textsc{Dustywave}: topological charges}
\label{app:EP3s_charges}

\begin{figure}
    \centering
    \includegraphics[width=\columnwidth]{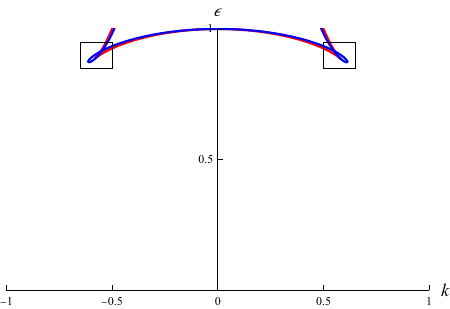}
    \includegraphics[width=\columnwidth]{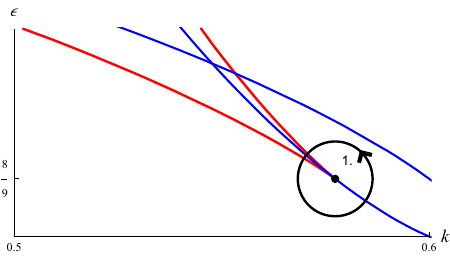}
    \includegraphics[width=\columnwidth]{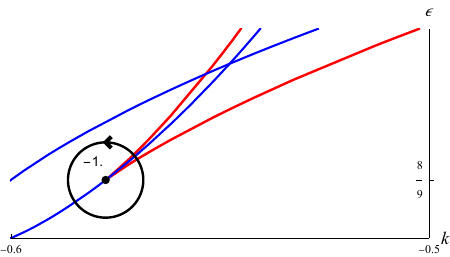}
    \caption{Exceptional Points of the \textsc{dustywave} in the parameter space ($k,\epsilon$). Intersections between the red ($R_1=0$) and blue ($R_2=0$) curves indicate positions of EP3s. Two regions are zoomed in on the two bottom panels. The EP3s are marked by black points, with associated topological charge of $\pm 1$.}
    \label{fig:dusty_EP3s}
\end{figure}

In addition to EPs of order 2, the \textsc{Dustywave} is also found to have EPs of order 3. The structure of these EP3s is folds of dimension 0, which are points in the ($k,\epsilon$) space. These points are located by a procedure which generalises the determination of EP2s. Noting $R_1 = \mathcal{R}(P,P')$ and $R_2 = \mathcal{R}(P,P'')$ the resultants of the characteristic polynomial and its derivatives, EP3s corresponds to points in the parameters space for which $P$,$P'$ and $P''$ have a common root. These points are those where both $R_1$ and $R_2$ cancel simultaneously. For the \textsc{Dustywave} problem, $R_{1}$ is given by Eq.~\eqref{eq:R1}, and $R_{2}$ by
\begin{equation}
    R_2 = 400 k^8 - 432 k^4 ( \epsilon-1 ) + 
 96 k^6 (14 - 51 \epsilon + 36 \epsilon^2).
\end{equation}
Fig.\ref{fig:dusty_EP3s} shows the curves $R_1=0$ (red) and $R_2=0$ (blue). These curves cross four times, and two of these points are EP3s. They are located at
\begin{equation}
    k = \pm \frac{1}{\sqrt{3}}, \quad \epsilon = \frac{8}{9}.
\end{equation}
For the other two points, $P,P''$ share a root different from the one shared by $P,P'$. As such, they are EP2s only. At the EP3s, $H$ has two eigenvalues, 0 and $\frac{-i}{3}$. It is non-diagonalizable, since it only has two eigenvectors,
\begin{eqnarray}
    \begin{pmatrix}
        0 \\ 1 \\ 0 \\ 0
    \end{pmatrix},
    \begin{pmatrix}
        -i \sqrt{3}/2 \\ 2 i \sqrt{3} \\ -1/2 \\ 2
    \end{pmatrix}.
\end{eqnarray}
The EP3s have a non-zero topological charge, as defined in \citep{delplace2021}. These topological invariants are defined as the winding numbers
\begin{equation}
    W_3 = \frac{1}{2\pi}\oint_C \frac{1}{R_1^2+R_2^2}(R_1\nabla_{\bs{p}} R_2 - R_2\nabla_{\bs{p}} R_1)\mathrm{d}\bs{p} ,
\end{equation}
for $\nabla_{\bs{p}}  = \begin{pmatrix}\partial_k \\ \partial_\epsilon\end{pmatrix}$ and $\bs{p}  = \begin{pmatrix}k \\ \epsilon\end{pmatrix} \in C$ where $C$ is any closed loop around one of the EP3s. We compute 
\begin{eqnarray}
    W_3\Big|_{\mathrm{EP3}_1} &=& +1,\\
    W_3\Big|_{\mathrm{EP3}_2} &=& -1,
\end{eqnarray}
where $\mathrm{EP3}_1 $ is the point at $(k,\epsilon) = (+\frac{1}{\sqrt{3}},\frac{8}{9})$ and $\mathrm{EP3}_2 $ is the point at $(k,\epsilon) = (-\frac{1}{\sqrt{3}},\frac{8}{9})$.

\section{\textsc{Dustywave}: \texorpdfstring{Anti-$\mathcal{PT}$}x operator }
\label{app:CP_operator}

Define $\Theta$ the operator of complex conjugation and $\mathcal{CP} \equiv \Tilde{U}\Theta$ the operator for anti-$\mathcal{PT}$ symmetry. Equation~\eqref{eq:CP} can be written under the form
\begin{equation}
    H_{\rm dw}\mathcal{CP}+\mathcal{CP}H_{\rm dw}=0 .
    \label{eq:commutCP_dusty}
\end{equation}
Eq.~\eqref{eq:commutCP_dusty} constrains the phase between the components of the eigenvectors of $H_{\rm dw}$, also called \textit{relations of polarisation}. Consider $X_0$ an eigenvector of $H_{\rm dw}$ with pure imaginary eigenvalue. Formally,
\begin{equation}
H_{\rm dw}X_0 = i\eta_0 X_0 ,
\label{eq:eigenval_dw}
\end{equation}
with $\eta_{0}$ real. Applying the operator $\mathcal{CP}$ on Eq.~\eqref{eq:eigenval_dw} and the commutation rule Eq.\eqref{eq:commutCP_dusty} gives the relation 
\begin{equation}
    H_{\rm dw} (\mathcal{CP} X_0) = i\eta_0 (\mathcal{CP} X_0).
\end{equation}
If the eigenvalue is not degenerated (i.e. the parameters are chosen for the system not to be at an Exceptional Point), $\mathcal{CP}X_0$ is colinear to $X_0$, i.e. 
\begin{equation}
    \mathcal{CP}X_0 = \alpha X_0
    \label{eq:prop}
\end{equation}
for some complex number $\alpha$. Hence, $H_{\rm dw}$ and $\mathcal{CP}$ share the same eigenvectors and the anti-$\mathcal{PT}$ symmetry is said to be {\it unbroken}. As a remark, the same reasoning on the operator $iH_{\rm dw}$ would lead to show that $iH_{\rm dw}$ is $\mathcal{PT}$-symmetric. $\mathcal{CP}$ can be straightforwardly diagonalised. Its eigenvectors are
\begin{equation}
    \begin{pmatrix}
        1 \\ 0 \\ 0 \\ 0
    \end{pmatrix},
    \begin{pmatrix}
        0 \\ 1 \\ 0 \\ 0
    \end{pmatrix},
    \begin{pmatrix}
        0 \\ 0 \\ i \\ 0
    \end{pmatrix},
    \begin{pmatrix}
        0 \\ 0 \\ 0 \\ i
    \end{pmatrix}.
\end{equation}
The condition given by Eq.~\eqref{eq:prop} imposes therefore the particular polarisation relations on non-propagating waves:
\begin{eqnarray}
    \left(\frac{\rho^\prime_\gas}{\rho^\prime_\dust}\right)^* &=& +\frac{\rho^\prime_\gas}{\rho^\prime_\dust}, \label{eq:unbrokenPol1}\\
    \left(\frac{\rho^\prime_\gas}{v^\prime_\gas}\right)^* &=& -\frac{\rho^\prime_\gas}{v^\prime_\gas}, \\
    \left(\frac{\rho^\prime_\dust}{v^\prime_\dust}\right)^* &=& -\frac{\rho^\prime_\dust}{v^\prime_\dust},\label{eq:unbrokenPol3}
\end{eqnarray}
and other linear combinations of Eqs.~\ref{eq:unbrokenPol1} -- \ref{eq:unbrokenPol3}. Hence, the analysis of the discrete symmetries of the problem shows with a very few amount of calculations that gas and dust densities are in phase as well as their velocities, while densities and velocities are in phase quadrature. Similar relations exist between fields in the unbroken $\mathcal{PT}$ symmetry phase of the Kelvin-Helmholtz problem discussed by \cite{fu2020}.

This result can be reformulated with real fields. In the unbroken anti-$\mathcal{PT}$ symmetry phase, the perturbations are damped without oscillating, and are of the form
\begin{eqnarray}
    \rho^\prime_\gas &\propto& \mathrm{e}^{-\eta_0 t}\cos{(kx)},\\
    \rho^\prime_\dust &\propto& \mathrm{e}^{-\eta_0 t}\cos{(kx)},\\
    v^\prime_\gas &\propto& \mathrm{e}^{-\eta_0 t}\sin{(kx)},\\
    v^\prime_\dust &\propto& \mathrm{e}^{-\eta_0 t}\sin{(kx)},
\end{eqnarray}
which imposes for quantities to be zero when they are spatially averaged. Formally, with
\begin{equation}
    \langle f \rangle (t) \equiv \lim_{L\to \infty} \frac{1}{L} \int_{-L/2}^{L/2} f(x,t)\mathrm{d}x,
\end{equation}
one obtains for example
\begin{eqnarray}
     \langle v^\prime_\gas \rho^\prime_\gas \rangle  &=& 0, \label{eq:avg1}\\
     \langle v^\prime_\dust \rho^\prime_\dust \rangle  &=& 0,\label{eq:avg2}
\end{eqnarray}
for all times. On the contrary, in the phase where anti-$\mathcal{PT}$ symmetry is broken, the frequency of the perturbations have a real part. Waves are propagating, and the polarisation relations are different than Eqs.\eqref{eq:unbrokenPol1}-\eqref{eq:unbrokenPol3}, and time-average of quantities defined in Eqs.~\ref{eq:avg1} -- \ref{eq:avg2} are non zero for this phase. Only propagating waves have non-zero macroscopic mass flux.

\end{document}